\documentclass[fleqn,usenatbib]{mnras}

\usepackage{amsmath}
\usepackage{txfonts}                
\usepackage{bm}                     
\usepackage{graphicx}               
\usepackage[T1]{fontenc}
\hypersetup{linkcolor=red}          

\newcommand{\convolution}{\mathop{\scalebox{1.5}{\raisebox{-0.2ex}{$\ast$}}}}
\newcommand{\kms}{\hbox{km s$^{-1}$}}
\newcommand{\dd}{\mathrm{d}}
\newcommand{\ppxf}{\textsc{ppxf}}
\newcommand{\norm}[1]{\left\lVert #1 \right\rVert}

\def\equationautorefname~#1\null{equation~(#1)\null}

\title[Improving full spectrum fitting]{Improving the full spectrum fitting method: accurate convolution with Gauss-Hermite functions}

\author[M.~Cappellari]{Michele Cappellari\thanks{E-mail:
michele.cappellari@physics.ox.ac.uk}\\
Sub-Department of Astrophysics, Department of Physics, University of Oxford, Denys Wilkinson Building, Keble Road, Oxford, OX1 3RH}

\date{Accepted 2016 November 18. Received 2016 November 18; in original form 2016 July 28}

\pagerange{\pageref{firstpage}--\pageref{lastpage}} \pubyear{2016}

\begin{document}
\label{firstpage}
\maketitle

\begin{abstract}
I start by providing an updated summary of the penalized pixel-fitting (\ppxf) method, which is used to extract the stellar and gas kinematics, as well as the stellar population of galaxies, via full spectrum fitting. I then focus on the problem of extracting the kinematic when the velocity dispersion $\sigma$ is smaller than the velocity sampling $\Delta V$, which is generally, by design, close to the instrumental dispersion $\sigma_{\rm inst}$. The standard approach consists of convolving templates with a discretized kernel, while fitting for its parameters. This is obviously very inaccurate when $\sigma\la\Delta V/2$, due to undersampling. Oversampling can prevent this, but it has drawbacks. Here I present a more accurate and efficient alternative. It avoids the evaluation of the under-sampled kernel, and instead directly computes its well-sampled analytic Fourier transform, for use with the convolution theorem. A simple analytic transform exists when the kernel is described by the popular Gauss-Hermite parametrization (which includes the Gaussian as special case) for the line-of-sight velocity distribution. I describe how this idea was implemented in a significant upgrade to the publicly available \ppxf\ software. The key advantage of the new approach is that it provides accurate velocities regardless of $\sigma$. This is important e.g. for spectroscopic surveys targeting galaxies with $\sigma\ll\sigma_{\rm inst}$, for galaxy redshift determinations, or for measuring line-of-sight velocities of individual stars. The proposed method could also be used to fix Gaussian convolution algorithms used in today's popular software packages.
\end{abstract}


\section{Introduction}

The kinematics of the stars and gas in galaxies is a key ingredient in our understanding of how they form and evolve. Nowadays, this information is routinely extracted from integral-field spectroscopic (IFS) data, which provide a three dimensional view of galaxies. The technology is now available on all major telescopes and has become the standard way of obtaining spectra for galaxy evolution studies \citep[see][for a review]{Cappellari2016}.

The IFS data provide a ``fossil record'' of galaxy formation. Specifically, the IFS kinematics of the stars allow one to distinguish a galaxy assembly dominated by gas poor merging from a growth driven by gas accretion and star formation \citep[e.g.][]{Emsellem2011,Cappellari2013p20,Naab2014}. Dynamical models based on IFS kinematics allow one to infer galaxy mass distributions to study scaling relations \citep[e.g.][]{Cappellari2013p15,Scott2015}, the stellar and dark matter content \citep[e.g.][]{Cappellari2012,Cappellari2015dm}, or measure black hole masses \citep[e.g.][]{Krajnovic2009,Seth2014,Walsh2016}. The gas content and kinematics tell us about the role of gas accretion in galaxy assembly \citep[e.g.][]{Sarzi2006,Davis2011b,Barrera-Ballesteros2015} or the mechanism that regulates star formation in galaxies \citep[e.g.][]{Alatalo2011,Cheung2016,Ho2016}.

For more than a decade, this information has been extracted from IFS data by surveys targeting one galaxy at a time (SAURON \citealt{deZeeuw2002}; DiskMass \citealt{Bershady2010}; ATLAS$^{\rm 3D}$ \citealt{Cappellari2011a}; CALIFA \citealt{Sanchez2012}). But the observational panorama is undergoing a revolution, with the arrival of large multiplexed IFS surveys targeting 10--20 galaxies at a time. Thousands of galaxies have already been observed in this manner (MaNGA \citealt{Bundy2015}; SAMI \citealt{Bryant2015}).

These ongoing IFS surveys depend critically on the full spectrum fitting technique to deliver their science. The method is used to extract stellar and gas kinematics, as well as stellar population from the spectral data cubes. But, given the large number of objects and the variety of galaxy morphological types, these surveys are pushing the existing techniques to their limits.

This paper is motivated by the existence of these ongoing large surveys and mostly arises from initial experiences with the analysis of the MaNGA data. In fact, one new characteristic of both the MaNGA and SAMI survey is that they are observing large numbers of galaxies (especially spirals) with stellar velocity dispersion $\sigma$ well below the instrumental dispersion $\sigma_{\rm inst}$ of the spectrographs. This situation was until now not very common, as the observers generally tried to target galaxies with $\sigma_{\rm inst}\la\sigma$. However this restriction is actually not necessary to obtain useful kinematic information. In fact, although it is true that $\sigma$ becomes intrinsically difficult to measure reliably when $\sigma\ll\sigma_{\rm inst}$, the kinematics of low-$\sigma$ galaxies is dominated by the stellar velocity $V$ \citep[e.g.][]{Cappellari2016}, which remains a well defined observable. I realized that in this observational regime, all the spectrum fitting approaches, which are nearly universally adopted to extract stellar and gas kinematics, suffer from limitations and can be significantly improved. Here in \autoref{sec:concepts} I describe general concepts about kinematic extraction, in \autoref{sec:ppxf} I give an updated summary of the \ppxf\ method, in \autoref{sec:low_dispersion} I discuss problems of the current approach and propose a clean solution, I summarize my paper in \autoref{sec:summary}.

\section{General concepts}
\label{sec:concepts}

For external galaxies, or other stellar systems not resolved into individual stars, the stellar kinematics information is extracted by comparing the observed spectra, at different spatial locations, with broadened stellar templates. This comparison is nowadays almost invariably performed in pixel space rather than in Fourier space \citep[e.g.][]{Rix1992losvd,vanDerMarel1994m87,Kelson2000, Cappellari2004}. This is because it makes the masking of bad pixels, or the simultaneous extraction of gas kinematics and stellar population, an easy and accurate process, while also allowing for other observational effects to be easily modelled.

In this section I review some general concepts, useful for understanding how the kinematics is generally extracted from galaxy spectra. I also try to clarify aspects that I discovered can lead to potential mistakes, from more than a decade of feedback I received from users of my publicly-available Penalized Pixel-Fitting software \ppxf\ \citep{Cappellari2004}.

\subsection{Templates observed like the galaxy spectrum}
\label{sec:log_rebin}

More than a decade ago, the stellar templates were generally observed with the same spectrograph used to obtain the galaxy spectrum under analysis \citep[e.g.][]{vanderMarel1994}. In this way one could assume that all the difference between the galaxy spectrum and the templates was due to the broadening by the galaxy line-of-sight velocity distribution (LOSVD). In this idealized case one could schematically model the galaxy spectrum $G_{\rm mod}$ as follows
\begin{equation}\label{eq:t_conv_b}
G_{\rm mod}(x)=T(x)\convolution \mathcal{L}(c x),
\end{equation}
where $x=\ln\lambda$ is the natural logarithm of wavelength, $T(x)$ is the stellar template, $\mathcal{L}$ is the LOSVD, with $c$ the speed of light and $\convolution$ denoting convolution.

It is worth noting that the use of a logarithmic axis, although universally adopted in the past few decades, is not strictly required for the kinematic extraction. It ensures a constant velocity scale for every $x$. This means that $\mathcal{L}(v)$ remains constant along the $x$ coordinate and allows one to use efficient Fast Fourier Transform (FFT) methods when performing the convolution. 

In principle however, it would also be possible to perform the convolution in linear $\lambda$. For this one would need to allow for the $\mathcal{L}(v)$ to vary with $x$, but one could still construct a relatively efficient convolution, via direct summation, by making use of the fact that the convolution kernel $\mathcal{L}(v)$ generally consists of a few orders of magnitude fewer elements than the galaxy and templates spectra. However, given that at least one resampling is unavoidable when reducing the spectral data, there are currently no compelling reasons to stop using logarithmic coordinates.

\subsection{Templates from spectral libraries}
\label{sec:templates_library}

In the past decade, thanks to the availability of high resolution empirical stellar libraries spanning large ranges of atmospheric parameters over a wide wavelength range \citep[e.g.][]{Cenarro2001,Prugniel2001,Valdes2004,Sanchez-Blazquez2006}, libraries of synthetic stellar spectra \cite[e.g.][]{Coelho2005,Munari2005,Gustafsson2008}, and stellar population models based on these libraries \citep[e.g.][]{bruzual03,Vazdekis2010,Maraston2011,Conroy2012models},  it has become common practice to extract the stellar kinematics by employing as templates linear combinations of a large number (up to a thousand) of template spectra observed with a different spectrograph than the one used to obtain the galaxy spectra. 

Due to this difference, before the kinematic extraction, and assuming the templates have a better resolution that the galaxy spectrum, the templates need to be matched to the galaxy spectrum by convolving them with a kernel $K(x,\lambda)$, which generally {\em varies} with wavelength $\lambda$, defined by the relation
\begin{equation}\label{eq:lsf_conv_k}
{\rm LSF}_{\rm inst}(x,\lambda) = {\rm LSF}_{\rm temp}(x,\lambda) \convolution K(x,\lambda).
\end{equation}
Here ${\rm LSF}_{\rm inst}(x,\lambda)$ is the line-spread function of the spectrograph used to observe the galaxy under study, which describes the observed shape of an intrinsically very narrow spectral line (in principle a $\delta$ function) at the wavelength $\lambda$, due to purely instrumental broadening, including the effect of integrating over the detector pixels. ${\rm LSF}_{\rm temp}(x,\lambda)$ is the LSF of the spectrograph used to observe the templates. After the templates have been convolved as
\begin{equation}\label{eq:t_conv_k}
\tilde{T}(x) = T(x) \convolution K(x,\lambda),
\end{equation}
they become as if they had been observed with the same spectrograph as the galaxy, and one can still model the galaxy spectrum with \autoref{eq:t_conv_b}. Note that \autoref{eq:t_conv_k} is not a standard convolution, due to the $\lambda$ dependence in the kernel. In signal processing this is called general convolution or general superposition integral.

In principle the kernel $K(x,\lambda)$ could be measured in the very same way one uses to extract the LOSVD from galaxies, using solar templates to fit spectra of the twilight sky, while allowing for a non-Gaussian shape for $K(x,\lambda)$. However, in practice, it is quite difficult to accurately characterize the spectral, spatial and temporal variation of both the ${\rm LSF}_{\rm inst}$ and ${\rm LSF}_{\rm temp}$ in the data. Moreover, the LSFs turn out to be generally well represented by Gaussian functions and, for this reason, are generally assumed to be Gaussians.

It is well know that the convolution of two normalized Gaussians with dispersion $\sigma_a$ and $\sigma_b$ is another normalized Gaussian with dispersion $\sigma_c^2=\sigma_a^2+\sigma_b^2$ \cite[e.g.][\S 2]{brigham1974fft}.
For these reasons, when the LSF are assumed Gaussian, the kernel in \autoref{eq:t_conv_k} is a normalized Gaussian
\begin{equation}\label{eq:gauss_diff}
G_{\rm diff}(x,\lambda)=\frac{\exp\left\{-x^2/\left[2\,\sigma_{\rm diff}^2(\lambda)\right]\right\}}{\sigma_{\rm diff}(\lambda)\sqrt{2\pi}}
\end{equation}
with dispersion 
\begin{equation}\label{eq:sigma_diff}
\sigma_{\rm diff}^2(\lambda)=\sigma_{\rm inst}^2(\lambda)-\sigma_{\rm temp}^2(\lambda),
\end{equation}
where $\sigma_{\rm inst}$ and $\sigma_{\rm temp}$ are the Gaussian dispersion of the LSF for the instrument used to observe the galaxy and for the one used to obtain the stellar templates, respectively.

Given the $\lambda$ dependence of $\sigma_{\rm diff}$, the general convolution in \autoref{eq:t_conv_k} needs to be performed via direct summation rather than using FFTs. But this only need to be done once, before fitting the LOSVD and for this reason its computation is not time critical. However, special care needs to be taken to avoid under-sampling during the computation of \autoref{eq:t_conv_k}, as it can be affected by the same under-sampling issues discussed in \autoref{sec:problem}.

\subsection{From measured velocity to observed redshift}

One important aspect that often causes confusion to users of spectral fitting software, is the connection between the measured line-of-sight velocity $V$ and the galaxy redshift $z$. For this reason I clarify this aspect here. 

As described in \autoref{sec:log_rebin}, spectral fitting codes rebin the spectra logarithmically and measure velocities from the shift $\Delta x$ required to match the spectral templates to the galaxy spectrum. This shift is precisely related to the redshift as follows
\begin{equation}\label{eq:pixel_shift}
\Delta x=\Delta\ln\lambda=\ln(\lambda_{\rm obsv}/\lambda_{\rm emit})=\ln(1+z).
\end{equation}

However, spectral fitting codes also need to define a velocity scale to be used as variable in the LOSVD (e.g.\ to quantify the galaxy velocity dispersion). For small velocity differences $V\equiv\Delta V$, the Doppler formula is
\begin{equation}\label{eq:doppler}
V\approx c\frac{\Delta\lambda}{\lambda_{\rm emit}}\approx c\, \Delta\ln\lambda.
\end{equation}
So one can {\em define} a constant velocity scale per pixel, which reduces to the Doppler formula in the limit of small $V$, namely on the scale of the LOSVD
\begin{equation}\label{eq:velscale}
V\equiv c\,\Delta x = c\, \Delta \ln\lambda = c\, \ln(1+z).
\end{equation}
This definition is used to convert the pixels differences $\Delta x$ measured by the codes into velocities, to provide the $V$ and $\sigma$ in \kms.
Given this definition, one has to use \autoref{eq:velscale} to convert back the measured velocities into redshifts, when needed. In particular, the well known approximation $z\approx V/c$  should {\em never} be used, being already less accurate than the typical measurement uncertainties when $z\ga0.01$.

\subsection{Separating peculiar velocities and cosmological redshift}

In the common case of IFS data, one is interested in the stellar and gas velocities $V_{\rm bin}$ measured within different spatial bins with respect to the galaxy barycentre (e.g.\ to construct dynamical models, or measure gas outflows) and not in the actual redshift $z_{\rm bin}$ of the individual spectra. 
When the redshift is negligible ($z\la0.01$), the $V_{\rm bin}$ can be computed with sufficient accuracy by simply subtracting the velocity $V_{\rm syst}$ of the galaxy barycentre from the measured velocities $V$ directly provided by the spectral fitting program. However this simple approach will lead to {\em dramatic} errors at larger redshift, where a quite different approach must be used.

In cosmology, the $V_{\rm bin}$ are called `peculiar velocities', to distinguish them from the cosmological recession velocity or redshift $z_{\rm cosm}$. The latter is formally defined as the redshift of the galaxy barycentre. For galaxies at significant redshift, an accurate and simple approach to measure $V_{\rm bin}$ consists of bringing all the galaxy spectra onto the rest frame wavelength $\lambda_{\rm rest}=\lambda_{\rm obs}/(1+z'_{\rm cosm})$, where $\lambda_{\rm obs}$ is the observed wavelength and $z'_{\rm cosm}$ is an initial estimate of the galaxy redshift. This division is actually equivalent to merely re-defining the origin of the velocity coordinate, while keeping the spectra unchanged. Crucially, the galaxy ${\rm LSF}_{\rm inst}(x,\lambda)$ must be compressed by the same factor or, when this is assumed Gaussian, its instrumental dispersion in units of wavelength $\sigma_{\rm inst}^{\rm rest}=\sigma_{\rm inst}^{\rm obs}/(1+z'_{\rm cosm})$. When quantified as $R=\Delta\lambda/\lambda$, in \kms, or as $\Delta\ln\lambda$, the instrumental resolution is independent of redshift, for every spectral pixel, but the pixels wavelength changes. This implies that, in general, the matching of the templates resolution of \autoref{eq:sigma_diff} must still be performed for every different redshift.

Once these two steps have been performed, the velocities $V$ returned by the spectral fitting program will be the desired $V_{\rm bin}$. From the extracted rest-frame velocity field, one can accurately measure possible residual offsets $V_{\rm bary}$ of the barycentre velocity from zero, by enforcing symmetries in the field \citep[e.g.][appendix~C]{Krajnovic2006}. This {\em small} offset can just be subtracted from all $V_{\rm bin}$. Finally, an improved estimate of the galaxy redshift can be obtained using the relation \citep[e.g.][eq.~2]{Cappellari2009}
\begin{equation}
1+z_{\rm cosm} = (1+z'_{\rm cosm})\times(1+V_{\rm bary}/c).
\end{equation}
which derives form the general expression linking the cosmological redshift $z_{\rm cosm}$, the bins peculiar velocities $V_{\rm bin}$ and the observed bins redshift $z_{\rm bin}$ \citep[e.g.][eq.~10]{Hogg1999}
\begin{equation}
1+z_{\rm bin}=(1+z_{\rm cosm})\times(1+V_{\rm bin}/c).
\end{equation}

\section{Overview of the pPXF method}
\label{sec:ppxf}

One particular technique used to extract stellar kinematics is called the Penalized Pixel-Fitting (\ppxf) method. It was originally described in \cite{Cappellari2004}, but it has significantly evolved over more than a decade of intense usage, to address our own specific needs and requests or feedback from users. A number of the new features in \ppxf\ were not described in detail in the literature, this is true in particular for the use of the covariance matrix (\autoref{sec:linear}), the implementation of regularization, for stellar population studies (\autoref{sec:regularization}), the fitting of gas emission lines (\autoref{sec:gas}) and the kinematic bulge-disk decomposition (\autoref{sec:bugle_disk}). For this reason, both the new and old features are summarized in this section. I describe here the current version 6.0 of the Python implementation of the software\footnote{Available from \url{http://purl.org/cappellari/software}}. But most of the features are also implemented in the IDL version of \ppxf.

\subsection{Model for the galaxy spectrum}

The \ppxf\ method approximates the observed galaxy spectrum via the following quite general parametrization
\begin{align}\label{eq:model}
G_{\rm mod}(x) & =
\sum_{n=1}^{N} w_n \left\{\left[T_n(x) \convolution \mathcal{L}_n(c x)\right]
\sum_{k=1}^{K} a_k \mathcal{P}_k(x)\right\} \nonumber\\
 &\quad + \sum_{l=0}^{L} b_l \mathcal{P}_l(x) 
 + \sum_{j=1}^{J} c_j S_j(x),
\end{align}
where the $\mathcal{L}_n$ are the LOSVDs, which can be different for the $N$ templates $T_n$ \cite[e.g.][]{Johnston2013}, the $\mathcal{P}_k$ and $\mathcal{P}_l$ are multiplicative or additive orthogonal polynomials of degree $k$ and $l$ respectively (of Legendre type or a truncated Fourier series), and $S_j$ are spectra of the sky \cite[e.g.][]{Weijmans2009}. The additive polynomials can significantly minimize template mismatch by changing the strength of individual absorption lines. Mismatch can be non negligible even when using today's large stellar libraries as templates. Additive polynomials can additional correct for imperfect sky subtraction or scattered light. The multiplicative polynomials can correct for inaccuracies in the spectral calibration, and make the fit insensitive to reddening by dust, without the need to adopt a specific reddening curve. Note that the 0th degree multiplicative polynomial is not included. However, if the calibration is reliable and one is interested in estimating the reddening from the shape of the stellar continuum, the multiplicative polynomials $\mathcal{P}_k(x)$ can be replaced by the expression
\begin{equation}
f(x) = 10^{-0.4\, E(B-V)\, k(x)}
\end{equation}
where $k(x)$ is a specific reddening curve \citep[e.g.][]{Cardelli1989,Calzetti2000} and $E(B-V)$ quantifies the amount of reddening, which is constrained by \ppxf\ to be positive. Any of the polynomials or sky components is optional and does not need to be actually used during a \ppxf\ fit.

There is considerable freedom in the choice of templates depending on the specific application. The templates can consist of combinations of e.g.\ (i) stellar population models with specific parameters 
\citep[e.g.][]{bruzual03,Vazdekis2010,Maraston2011,Conroy2012models}; (ii) individual empirical
\cite[e.g.][]{Cenarro2001,Prugniel2001,Valdes2004,Sanchez-Blazquez2006} or synthetic \cite[e.g.][]{Coelho2005,Munari2005,Gustafsson2008} stars;
(iii) principal components derived from a library of spectra; (iv) weighted sum of different stars; (v) gas emission lines; or (vi) sky spectra. 
The first option is used when \ppxf\ is employed for `full spectrum' fitting to study the galaxies stellar population \citep[e.g.][]{Onodera2012,Cappellari2012,Morelli2013,Morelli2015,McDermid2015,Shetty2015}, but it can also be used when extracting the stellar kinematics with \ppxf\ \citep[e.g.][]{Emsellem2004,Oh2011,Thomas2013boss}, while the remaining choices are useful when extracting the stellar or gas kinematics with \ppxf, due to the extra freedom which allows one e.g. to fit possible variations in elemental abundances \citep[e.g.][]{Cappellari2011a,Blanc2013,vanDeSande2016}.

\subsection{Parametrization for the LOSVD}

The LOSVD $\mathcal{L}_n(v)$, for both the stellar and gas templates, are parametrized using the Gauss-Hermite parametrization introduced for this purpose by \cite{vanDerMarel93} and \cite{Gerhard1993}. Note however that the two papers did not quite define the same parametrization for the LOSVD. The crucial difference is that, while \cite{Gerhard1993} fitted all moments in the Gauss-Hermite expansion, \cite{vanDerMarel93} chose to explicitly set the first three coefficients to $(h_0,h_1,h_2)=(1,0,0)$, and only fit the higher coefficients, in such a way that the LOSVD has the form
\begin{align}\label{eq:losvd}
\mathcal{L}(y)&=\frac{\exp\left(-y^2/2\right)}{\sigma\sqrt{2\pi}}
\left[ 1 + \sum_{m=3}^M h_m H_m(y) \right],\\
y&=(v-V)/\sigma \nonumber
\end{align}
It was the latter form, of \autoref{eq:losvd}, which has become the current standard in the field \cite[e.g.][]{Bender1994} and was also adopted by \ppxf.
Here, $H_m$ are the Hermite polynomials, standardized (by definition) in such a way that the terms of \autoref{eq:losvd} are quantum-mechanical wave functions for the one-dimensional harmonic oscillator \citep[e.g.][\S~4]{schiff1968quantum}
\begin{equation}
H_m(y)=\frac{H_m^{\rm Abr}(y)}{\sqrt{m!\,2^m}},
\end{equation}
were the $H_m^{\rm Abr}$ are the so-called `physicists' Hermite polynomials, defined as in \citet[eq.~22.2.14]{Abramowitz1964}. The $H_m^{\rm Abr}$ polynomials are the ones most commonly provided by default by today's popular software. They are e.g.\ the form returned by \textsc{Mathematica}'s \citep{wolfram2003mathematica} function \texttt{HermiteH}, or by \textsc{Numpy}'s \citep{Numpy2007} \texttt{polynomial.hermite.hermval}, or by \textsc{MATLAB}'s function \texttt{hermiteH}. The first two Hermite polynomials $H_m$ in \autoref{eq:losvd} are \citep[e.g.][eq.~A5]{vanDerMarel93}
\begin{equation}
H_3=\frac{y \left(2 y^2-3\right)}{\sqrt{3}},\qquad
H_4=\frac{4 \left(y^2-3\right) y^2+3}{\sqrt{24}}.
\end{equation}

\subsection{Linear fitting procedure}
\label{sec:linear}

For every choice for the possible non-linear parameters in the model of \autoref{eq:model}
\begin{equation}\label{eq:nonlinear_parameters}
 \left[V,\sigma,h_3,\ldots,h_M,a_1,\ldots,a_K,E(B-V)\right],
\end{equation} 
\ppxf\ minimizes the functional \citep[e.g.][\S 19.5]{Press2007}
\begin{equation}\label{eq:penalty_functional}
\mathcal{F} = \chi^2 + \lambda\mathcal{B} 
\end{equation}
where the first term measures the agreement between the model spectrum $G_{\rm mod}$ and the observed galaxy spectrum $G$, while the second is an adjustable term, which can be zero, and which quantifies the smoothness of the weights, in a space spanned by the population parameters (e.g. age, metallicity, $\alpha$ enhancement or the stellar initial mass function [IMF]). The computation of the first term of \autoref{eq:penalty_functional} is described in this section, while the \ppxf\ implementation of the second one is described in \autoref{sec:regularization}.

In the general case in which one knows the covariance matrix $\bm{\Sigma}$, with elements ${\rm cov}(x_j,x_k)$, were $x_j$ and $x_k$ are a pair of spectral pixels, the agreement between the data and the model, making the standard assumption of Gaussian uncertainties, is quantified by
\begin{align}\label{eq:chi2_matrix}
\chi^2 &= \left[\mathbf{A}\cdot\mathbf{x}-\mathbf{y}\right]^T \cdot \bm{\Sigma}^{-1} \cdot \left[\mathbf{A}\cdot\mathbf{x}-\mathbf{y}\right]\nonumber\\
&= \norm{\mathbf{L}^{-1}\cdot\left(\mathbf{A}\cdot\mathbf{x}-\mathbf{y}\right)}^2 \\
&= \norm{\mathbf{r}}^2,\nonumber
\end{align}
where $\norm{\cdot}$ is the Euclidean norm, the columns of the matrix $\mathbf{A}$ consist of the convolved templates (multiplied by the polynomials), the additive polynomials and the sky spectra, $\mathbf{y}=G$ is the galaxy spectrum,
\begin{equation}\label{eq:solution_vector}
\mathbf{x}=\left(w_1,\ldots,w_N,b_0,\ldots,b_L,c_1,\ldots,c_J\right)
\end{equation} 
is the solution vector, $\mathbf{L}=\sqrt{\mathbf{\Sigma}}$ is the square-root or the positive-definite covariance matrix $\bm{\Sigma}=\mathbf{L}\cdot\mathbf{L}^T$, computed via the Cholesky decomposition \citep[e.g.][\S 2.9]{Press2007}, and $\mathbf{r}$ is the weighted vector of residuals.

The minimum $\chi^2$ of \autoref{eq:chi2_matrix} is found by solving the following special quadratic programming problem
\begin{align}\label{eq:linear_system}
{\rm minimize}\quad & f(\mathbf{x})=\norm{\left(\mathbf{L}^{-1}\cdot\mathbf{A}\right)\cdot\mathbf{x}-\mathbf{L}^{-1}\cdot\mathbf{y}}^2\nonumber\\
{\rm subject\; to}\quad & w_n\ge0, \quad n=1,\ldots,N\\
& c_j\ge0, \quad j=1,\ldots,J\nonumber.
\end{align}
Specific, efficient and robust algorithms exist to solve this type of problems, with guaranteed convergence in a finite number of steps to the global minimum. Currently the IDL version of \ppxf\ uses the Bounded-Variables Least Squares (\textsc{bvls}) algorithm by \citet{Lawson1995}, while the Python version uses the Non-Negative Least Squares (\textsc{nnls}) code by the same authors, made available by the Scipy \citep{Scipy2001} procedure \texttt{optimize.nnls}. In the \textsc{nnls} solution, I employ slack variables to remove the positivity constraints from the additive polynomials coefficients $(b_0,\ldots,b_L)$.

In the common situation where spectral covariance is ignored, or unknown, and one only has the error spectrum, then 
\begin{equation}
\mathbf{L}^{-1}=\mathbf{diag}[1/\Delta G(x_1),\ldots,1/\Delta G(x_P)]
\end{equation}
reduces to a diagonal matrix, where $\Delta G(x_p)$ is the $1\sigma$ uncertainty of every pixel $x_p$ in the galaxy spectrum. In this special, but usual case, the computation can be simplified, as there is no need to decompose and multiply by the inverse covariance matrix, and the residual vector $\mathbf{r}$ has elements
\begin{equation}\label{eq:chi2_no_covar}
r_p = \frac{G_{\rm mod}(x_p)-G(x_p)}{\Delta G(x_p)}, \quad p=1,\ldots,P.
\end{equation}

\subsection{Non-linear fitting procedure}
\label{sec:non_linear}

A key feature of \ppxf, from which the method derives its name, is the fact that it automatically penalizes non-Gaussian solutions, to reduce the noise in the recovered kinematics, when the data do not contain sufficient information to constrain the full shape of the LOSVD. This is done by minimizing a new objective function
\begin{equation}\label{eq:penalty}
\chi^2_p=\chi^2 + \alpha\, \mathcal{P},
\end{equation}
where the $\chi^2$ is given in \autoref{eq:chi2_matrix} or \autoref{eq:chi2_no_covar}, $\mathcal{P}$ is a penalty function, which describes the deviation of the LOSVD from a Gaussian shape, while $\alpha$ is an adjustable penalty, which depends on the data quality. The penalty is implemented in an efficient way, by perturbing the residual vector $\mathbf{r}$ in such a way that the sum-of-squares nature of the problem is preserved. This is described in the original \ppxf\ paper \citep[section~3.3]{Cappellari2004} and will not be repeated here.

In both cases, with or without the covariance matrix, the minimization of the $\chi_p^2$ as a function of the non-linear parameters in  \autoref{eq:nonlinear_parameters} is performed with the specific Levenberg-Marquardt non-linear least-squares optimization algorithm \citep[e.g.][\S 15.5.2]{Press2007}. This exploits the sum-of-squares nature of the optimization problem by requiring the user to provide the vector of residuals $\mathbf{r}$, instead of the scalar $\chi^2$ of \autoref{eq:chi2_matrix}, to compute explicitly the Hessian matrix of the $\chi^2$ merit function. This makes the method much more robust and efficient than generic optimizers of scalar functions. 

\ppxf\ uses the state-of-the-art \textsc{minpack} implementation of the Levenberg-Marquardt method by \citet{More1980minpack}, which is based on the robust trust region method. It was converted into IDL and named \textsc{mpfit} by \citet{Markwardt2009}, with the important addition of the ability to set bounds or keep any of the variables fixed. \textsc{mpfit} was ported to Python by Mark Rivers and later adapted for use with Numpy by Sergey Koposov. 

In \ppxf, every individual non-linear parameter of the LOSVD, for every kinematic component separately, can be: (i) left free; (ii) kept fixed at a specified value; or (iii) bound within specified values. This flexibility has numerous applications. As an example, one can fit for the kinematic (e.g. $V$ and $\sigma$), templates and polynomials for the gas emission-line components in individual spaxels, while keeping the stellar kinematics (e.g. $V$, $\sigma$, $h_3$ and $h_4$) fixed to the value determined from a larger spatial bin.

A possible alternative for the non-linear optimization under Python would be the Scipy function \texttt{optimize.least\_squares}, introduced in version 0.18 in 2016, which is also designed for least-squares problems with bounds and is also based on the trust region method. I performed some tests using \ppxf\ with \textsc{least\_squares} for the non-linear optimization, trying to match as closely as possible both the accuracy requirement and the step size for the numerical derivatives, that I currently use in \ppxf\ with \textsc{mpfit}. I extracted the stellar and gas kinematics both from simulated spectra, with a range of $\sigma$, as well as from real MaNGA data cubes.
I found that \ppxf\ using \textsc{least\_squares} requires an execution time and a number of function evaluations which is generally comparable, but never smaller than the public \ppxf\ version using \textsc{mpfit}. This is true both when using the trust-region reflective algorithm (\texttt{method=`trf`; \citealt{branch1999subspace}}) and the dogleg algorithm (\texttt{method=`dogbox`}, \citealt{voglis2004dogleg}, \citealt[\S~4]{nocedal2006numerical}) in \textsc{least\_squares}. It was reassuring to see both \textsc{least\_squares} methods invariably converge to the same \textsc{mpfit} solution within the specified accuracy.  These tests indicates there is currently no compelling reason to replace \textsc{mpfit} in \ppxf.

\subsection{Weights regularization for stellar population}
\label{sec:regularization}

To study the star formation history and stellar population of a galaxy, \ppxf\ uses the common approach of modelling its unobscured rest-frame spectrum by discretizing the following integral equation \citep[e.g.][]{CidFernandes2005,Ocvirk2006,Tojeiro2007}
\begin{equation}\label{eq:sfh}
G_{\rm mod}(\lambda)=\int_{t=0}^{t=T} {\rm SSP}_\lambda(t,Z)\cdot {\rm SFR}(T-t)\,  \dd t,
\end{equation}
where SFR is the star formation rate, ${\rm SSP}_\lambda$ is a Single Stellar Population spectrum per unit mass, with age $t$ and metallicity $Z$, while $T$ is the age of the Universe at the redshift of the galaxy. This expression is straightforwardly generalized in \ppxf\ to study the distribution of more parameters, like e.g.\ metallicity, $\alpha$ enhancement or IMF, in addition to the SFR \citep[e.g.][\S 2.3]{Conroy2013}. 

The \autoref{eq:sfh} is an inhomogeneous Fredholm equation of the first kind, with kernel ${\rm SSP}_\lambda$. And the recovery of the ${\rm SFR}(t)$ from the observed $G_{\rm mod}$ is a textbook example of ill-conditioned inverse problem \citep[e.g.][\S 19.0]{Press2007}. This means that the recovery suffers from severe degeneracies and a unique solution cannot be found without further assumptions.
    
A standard way of dealing with this type of problem is by using regularization \citep[e.g.][]{tikhonov1977regul,hansen1998regul}, which can be thought of as damping the high-frequency variations in the solution, unless they are actually required to describe the data, or finding a trade-off between the quality of the fit and the noise in the solution  \citep[e.g.][\S 19.4]{Press2007}. In other words, regularization allows one to select the smoothest solution, among the many degenerate solutions that are {\em equally} consistent with the data. 

A conceptual mistake that is often made regarding regularization, is to think that this is equivalent to assuming the solution has to be smooth. This is {\em not} correct. In fact, the solution can be as non-smooth as required by the data and e.g.\ it will allow for multiple sharp bursts of star formation if the data require them. The solution will only be smooth if this is consistent with the data.

Under some simplifying assumptions, the regularized solution has a simple Bayesian interpretation: it represents the most likely solution for the weights, given an adjustable prior on the amplitude of the fluctuations \citep[e.g.][\S 19.4.1]{Press2007}. An alternative to regularization consists of explicitly exploring the posterior of the allowed solutions using a Markov Chain Monte Carlo approach \citep[e.g.][\S~11]{gelman2013bayesian}, while using the individual weights as non-linear model variables. Although potentially interesting and conceptually simple, we found the latter approach quite time consuming for general usage and did not yet include it in the public version of \ppxf.

Assuming, without loss of generality, that one is performing a spectral-fitting study of the stellar population, while varying a single parameter (e.g. age), the smoothness functional which appears in \autoref{eq:penalty_functional} is defined as follows \citep[e.g.][\S 19.5]{Press2007}
\begin{equation}\label{eq:regularization}
\lambda\mathcal{B} = \lambda\int w''(t)^2 \dd t\; \propto \sum_{n=2}^{N-1} \frac{\left(w_{n-1} - 2w_n + w_{n+1}\right)^2}{\Delta}
\end{equation}
where the $w_n$ are the spectral weights and the functional is zero only for a linear function of the weights. This expression is straightforwardly extended within \ppxf\ to dimensions large than one, by performing the finite differences separately along every axis. In this case the software requires the stellar population templates to form a full grid in 2-dim or 3-dim, in such a way that one can easily map the elements of the weights vector onto the 2-dim or 3-dim coordinates in population parameters. 

In practice, the regularization is implemented in \ppxf\ by augmenting the matrix in \autoref{eq:linear_system} with one extra row, namely one extra equation 
\begin{equation}\label{eq:differences}
\frac{w_{n-1} - 2w_n + w_{n+1}}{\Delta}=0,
\end{equation}
for every term in the summation of \autoref{eq:regularization}, in such a way that the regularized solution of the problem in \autoref{eq:linear_system} becomes
\begin{align}\label{eq:regul_minimization}
{\rm minimize} \quad & f(\mathbf{x})=\norm{\left(
\begin{array}{c}
\mathbf{L}^{-1}\cdot\mathbf{A}\\
\lambda\, \mathbf{C}
\end{array}
\right)
\cdot\mathbf{x}-
\left(
\begin{array}{c}
\mathbf{L}^{-1}\cdot\mathbf{y}\\
0
\end{array}
\right)}^2\nonumber\\
{\rm subject\; to}\quad & w_n\ge0, \quad n=1,\ldots,N\\
& c_j\ge0, \quad j=1,\ldots,J\nonumber,
\end{align}
with $\mathbf{C}$ the array with the coefficients of \autoref{eq:differences}. This augmented system is still solved by \textsc{bvls} or \textsc{nnls} as in the non-regularized case. Also here, the computation is specialized and simplified when the covariance matrix is not given. Applications of \ppxf\ with regularization to the study of the star formation history or stellar population in galaxies were presented e.g. in \citet{Onodera2012}, \citet{Cappellari2012}, \citet{Morelli2013,Morelli2015} and \citet{McDermid2015}.

A conceptually similar regularization approach was used in the \textsc{stecmap} full-spectrum fitting code \citep{Ocvirk2006}. However the \textsc{stecmap} implementation is very different. Their approach only regularizes in one dimension. Moreover it does not exploit the quadratic nature of the sub-problem of \autoref{eq:regul_minimization}, but instead solves it as a generic non-linear problem, using a variable-metric non-linear optimization method. This makes their approach less robust as well as much slower and more complex than the \textsc{nnls} approach used by \ppxf. Fast and accurate convergence of the solution of the quadratic sub-problem of \autoref{eq:regul_minimization} is essential for \ppxf\ to be able to reliably solve for the non-linear parameters of the LOSVD together with the template weights.

\subsection{Modelling gas emission lines}
\label{sec:gas}

The ability of \ppxf\ to fit different kinematics, or a different LOSVD $\mathcal{L}_n$, for every individual template also allows one to fit the gas emission lines together with the stellar kinematics and stellar population. This is achieved with \ppxf\ by simply passing a set of gas emission-lines templates, together with the stellar ones. In this way, \ppxf\ can fit for the kinematics and weights of both the stellar and gas components, without the need for the code to know, or make any algorithmic distinction, between the two sets of templates. However, given that the gas kinematics is often dominated by rotation rather than by random motions, the gas emission lines can have quite low dispersion. This implies that an accurate treatment of undersampling, which is the focus of this paper, is critical for unbiased gas measurements.

The generality of the \ppxf\ approach, where emission lines are just normal templates, gives flexibility to the users. One can freely specify the emission lines templates to include in the fit, and each line can be described by an arbitrary number of components if desired. The kinematics of every line can be tied to each other, or they can be fitted independently. And the relative fluxes of emission line doublets can be fixed, by placing two lines in the same template. The LOSVD of the gas emission lines are not restricted to being Gaussians, but they can be described by the same general Gauss-Hermite parametrization of \autoref{eq:losvd}, when needed, and in this case one can use the same penalty $\mathcal{P}$ of \autoref{eq:penalty}. Application of \ppxf\ to the simultaneous fit of the gas emission lines and the stellar population where presented e.g. in \citet{Johnston2013}, \citet{Shetty2015} and \citet{Mitzkus2016}.

It is worth mentioning that a simultaneous fit of both the stellar and gas kinematics should not be used as the standard approach. In general, more robust stellar kinematics is obtained by masking the gas emissions and including additive polynomials to reduce template mismatch. And the stellar kinematics is generally extracted from larger spatial bins than those needed for the more clumpy gas. Moreover, for extracting the gas fluxes and kinematics, as well as for the stellar population, one generally only includes multiplicative polynomials, to prevent changes in the line strength of the absorption features in the templates. For these reasons, in general, the gas kinematics is extracted from a separate \ppxf\ fit, at fixed stellar kinematics. The simultaneous fitting of gas and stellar templates is especially useful when studying stellar population, as one does not need to mask, or clean, regions of possible emission. This allows one to make full use e.g.\ of the spectral region of the Balmer series, where gas emission is common, and which provides strong constraints on the ${\rm SFR}(t)$.

In practice, the gas templates should represent the ${\rm LSF}_{\rm inst}(x,\lambda)$ at the wavelength of a given set of emission lines. Usually the LSF is approximated by a Gaussian, and for maximum accuracy, the gas templates need to be integrated over the spectral pixels, as is automatically the case for the stellar templates. Pixels integration is generally ignored by other software, however it has a small but measurable effect on the recovered gas dispersion as I will show in \autoref{sec:formula}. For a normalized (unit area) Gaussian LSF, the pixel-integrated gas template, for a single line of wavelength $\lambda_{\rm line}$, in the same logarithmic coordinates $x_p$ as the stellar templates, is
\begin{align}\label{eq:pixel_integration}
T_{\rm gas}(x_p)=& \frac{1}{2}\left\lbrace 
{\rm erf}\left[\frac{x_p-x_0+(\Delta x)_{\rm pixel}/2}{\sigma_{\rm inst}(x_0)\sqrt{2}}\right]\right.\nonumber \\
&- 
\left.{\rm erf}\left[\frac{x_p-x_0-(\Delta x)_{\rm pixel}/2}{\sigma_{\rm inst}(x_0)\sqrt{2}}\right]
\right\rbrace, \quad p=1,\ldots,P
\end{align}
where $x_0=\ln\lambda_{\rm line}$, $(\Delta x)_{\rm pixel}=(\Delta\ln\lambda)_{\rm pixel}$ is the size of one spectral pixel and erf is the error function \citep[eq.~7.1.1]{Abramowitz1964}. The gas template is band-limited as it is the convolution of a box function and a Gaussian with dispersion $\sigma_{\rm inst}$. When the template is Nyquist sampled at intervals $(\Delta x)_{\rm pixel}\approx\sigma_{\rm inst}$, it is uniquely defined by its samples and will result in accurate convolutions with the LOSVD to near machine accuracy. The Nyquist sampling is guaranteed by the fact that, if the galaxy spectra or templates were sampled at steps larger than $\sigma_{\rm inst}$, one would be {\em wasting} useful information in the data!

The \ppxf\ approach is closely related, but conceptually different from the one adopted by the popular and state-of-the-art \textsc{Gandalf} gas-fitting code \citep{Sarzi2006}, which was in fact derived from an earlier version of \ppxf, and first introduced the simultaneous fitting of multiple stellar and gas components, for the accurate subtraction of the stellar continuum. The \textsc{Gandalf} code treats the gas emission lines and the stellar templates in a separate manner: the non-convolved Gaussian emission lines are added to the stellar templates, which are convolved with a fixed stellar kinematics. Ultimately both the \ppxf\ and \textsc{Gandalf} approaches are expected to give comparable results for the gas emissions, when adopting similar settings. The key difference is that \ppxf\ was designed to extract stellar kinematics, for multiple components, and regularized stellar population, in addition to the gas kinematics.

\subsection{Kinematic bulge disk decomposition}
\label{sec:bugle_disk}

A recent addition to \ppxf\ is the possibility of constraining the ratio of the first two kinematic components to a desired value. This can be useful e.g.\ to perform kinematic bulge-disk decompositions \citep{Tabor2016}. One can force the template spectra for the bulge component to contribute a prescribed fraction $f_{\rm bulge}$ of the total flux in the fitted spectrum
\begin{equation}
f_{\rm bulge}=\frac{\sum w_{\rm bulge}}{\sum w_{\rm bulge} + \sum w_{\rm disk}},
\end{equation}
where $w_{\rm bulge}$ and $w_{\rm disk}$ are the weights assigned to the set of spectral templates used to fit the bulge and disk respectively.

This constraint is enforced in the same way as the regularization constraints of \autoref{eq:regul_minimization}, by adding the following extra equation to the matrix of \autoref{eq:linear_system}
\begin{equation}\label{eq:bulge_disk}
\frac{(f_{\rm bulge}-1)\sum w_{\rm bulge} + f_{\rm bulge}\sum w_{\rm disk}}{\Omega}=0,
\end{equation}
with the parameter $\Omega$ set to a very small number (e.g. $\Omega=10^{-9}$), which specifies the relative accuracy at which this equation needs to be satisfied. When both the spectral templates and the galaxy spectrum are normalized to have a mean flux of order unity, the best fitting weights $w_{\rm bulge}$ and $w_{\rm disk}$  are generally smaller than unity and \autoref{eq:bulge_disk} is satisfied as an equality constraint to numerical accuracy.

\begin{figure*}
    \includegraphics[width=\textwidth]{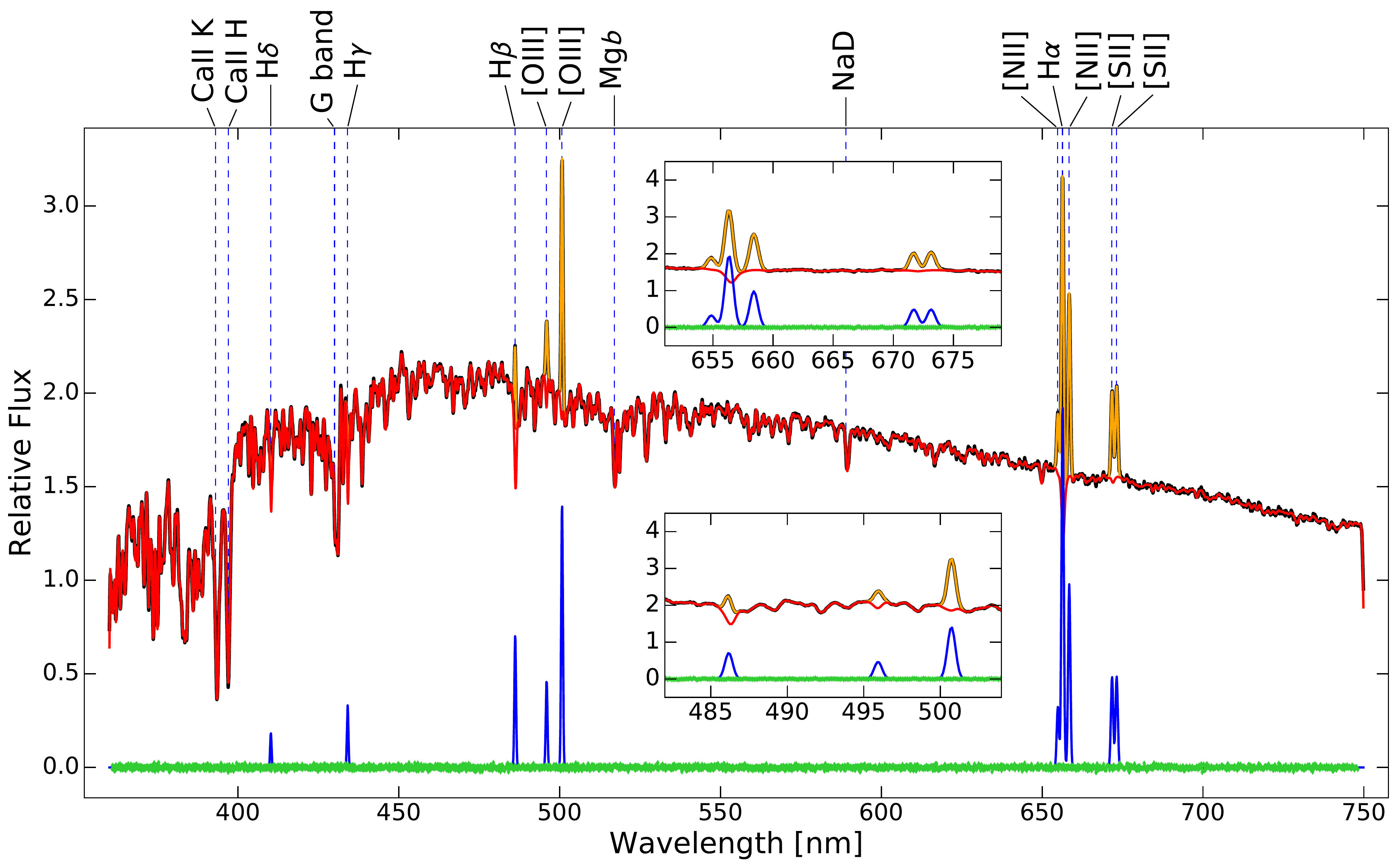}
    \caption{Solar spectrum with added gas emission lines. The figure illustrates a typical \ppxf\ fit to the solar spectrum by \citet{Kurucz2005}, and to the gas emissions, used for our tests. The black line (mostly hidden by the fit) is the relative flux of the observed spectrum (with noise added at $S/N=200$), for an adopted dispersion $\sigma_{\rm in}=140$ \kms, for both the gas and the stars. The red line is the \ppxf\ fit for the stellar component, while the orange line is a fit to the gas emission lines. The green symbols at the bottom are the fit residuals, while the blue lines is the gas-only best-fitting spectrum. The main absorption and emission features are indicated at the top of the plot. The two inset plots show an enlarged view of the \ppxf\ fits and residuals in the regions including the H$\alpha$ (top) and H$\beta$ (bottom) lines.
        \label{fig:solar_spectrum}}
\end{figure*}

\section{Extracting kinematics at low dispersion}
\label{sec:low_dispersion}

In this section I discuss the problem of extracting accurate kinematics using full spectrum fitting, in the special case in which the velocity dispersion is smaller than the velocity step $(\Delta V)_{\rm pixel}$. This is generally chosen to be similar to the instrumental dispersion $\sigma_{\rm inst}$. In fact, if $(\Delta V)_{\rm pixel}<\sigma_{\rm inst}$ the pixels have significant covariance and some data are redundant, while if $(\Delta V)_{\rm pixel}>\sigma_{\rm inst}$ the spectrum is not Nyquist sampled and one is wasting useful information. In what follows I assume $(\Delta V)_{\rm pixel}=\sigma_{\rm inst}$, and sometimes use the two interchangeably, when referring to the critical scale where issues appear.

I illustrate the problems and solution discussed in this section using numerical experiments. My goal is {\em not} to asses the general ability of \ppxf\ to recover galaxy kinematics, or other practical issues, like the biases as a function of the quality of the stellar templates. This has already been covered in the literature \citep[e.g.][]{Westfall2011}. Instead, I focus here exclusively on the important numerical problems due to the discretization  of the LOSVD.

\subsection{Setting up the numerical experiments}

For the experiments I restrict myself to the wavelength range 360--750 nm, which is a spectral region containing a number of useful absorption features for the kinematic extraction. This wavelength range is fully covered e.g. by the MaNGA \citep{Bundy2015} and CALIFA \citep{Sanchez2012} integral-field spectroscopic surveys. This is also the region covered by the spectra of the MILES stellar library \citep{Sanchez-Blazquez2006}. However the results I present are general and are quite insensitive to the adopted wavelength region. 

I use as my `galaxy' spectrum the solar spectrum\footnote{Available from \url{http://kurucz.harvard.edu/sun/}} by \citet{Kurucz2005}, which has a resolution $R=\Delta\lambda/\lambda\approx300\,000$ FWHM and a typical $S/N\approx3000$ per spectral element. This spectrum can be regarded as having essentially infinite resolution and $S/N$ for all practical purposes of my tests. Also in this case, I have verified that the results of the experiments are general and weakly depend on the specific choice of the input spectrum, as long as it is representative of real high-resolution galaxy spectra. Nearly indistinguishable results were obtained, in all my tests, using e.g.\ a high-resolution $R=20\,000$, solar metallicity, 10 Gyr, single stellar population model of \citet{Maraston2011}, based on the MARCS theoretical stellar library \citep{Gustafsson2008}.

To test the recovery of the gas kinematics, on top of the Solar spectrum, I add a realistic set of Gaussian gas emission lines. I include the Balmer series H$\alpha$, H$\beta$, H$\gamma$, H$\delta$, with relative fluxes following an unreddened Balmer decrement \citep{Storey1995}. I additionally include the [\ion{O}{iii}]~$\lambda\lambda$4959,5007 and [\ion{N}{ii}]~$\lambda\lambda$6548,6583 doublets, with fluxes in the standard ratio 1/3, and the [\ion{S}{ii}]~$\lambda\lambda$6716,6731 doublet with flux ratio of one. The doublets are scaled with respect to the Balmer lines according to the ratios [\ion{O}{iii}]~$\lambda5007/{\rm H}\beta=2$, [\ion{N}{ii}]~$\lambda6583/{\rm H}\alpha=1/2$ and [\ion{S}{ii}]~$\lambda\lambda6716,6731/{\rm H}\alpha=1/2$. These ratios are nearly irrelevant for the tests, but are chosen to be representative of realistic ionization levels for what \citet{Ho1997} calls ``transition objects'' between star-forming regions and LINERS. The Balmer series and the three doublets are all treated as separate kinematic components, in addition to the stellar one. This means that the \ppxf\ fits presented here have five kinematic components.

I adopt an idealized spectrograph, with a Gaussian LSF, with a constant instrumental dispersion $\sigma_{\rm inst}=70$  \kms\ (i.e. $R\approx1800$ FWHM). This is the median resolution of the MaNGA spectrograph, but is also similar to the one used by other galaxy surveys: it is similar e.g. to the resolution of the ATLAS$^{\rm 3D}$ survey \citep{Cappellari2011a}, to the high-resolution mode used by the CALIFA survey \citep{Sanchez2012}, and to the low-resolution mode used by the SAMI survey \citep{Bryant2015}. I assume the spectra are Nyquist sampled by the detector, and consequently adopt a spectral pixel $(\Delta x)_{\rm pixel} = \sigma_{\rm inst}=70$ \kms. This is the same velocity sampling adopted e.g.\ by the SDSS data release 13  \citep{sdss2016}.

I investigate the case where I use as input stellar template in \ppxf\ the same spectrum used for the galaxy spectrum. I further assume a quite high $S/N=200$ per spectral interval $(\Delta x)_{\rm pixel}$. Both choices are made again to provide a clean experiment which isolates the under-sampling problem from e.g. the unrelated issue of template mismatch. I adopt the default \texttt{degree=4} additive polynomials, but my results are totally insensitive to this choice. A representative \ppxf\ fit to the solar spectrum and to the added emission lines, is shown in \autoref{fig:solar_spectrum}.

In my experiments, the solar spectrum was initially logarithmically sampled with velocity scale of 2 \kms\ per pixel, which is an integer factor smaller than the final detector pixels. Then the gas emission lines were added. Subsequently, the whole spectrum was accurately convolved with a very well-sampled, discretized LOSVD and with a Gaussian LSF. Then the spectrum was integrated over the 70 \kms\ wide pixels, by summing every adjacent set of 35 pixels. Finally noise was added to every pixel. 
In the tests, the input velocity was chosen randomly for every realization, to prevent the LOSVD from being aligned in a constant manner with respect to the pixels boundaries. The same applies to the starting guess for $V$ and $\sigma$ in \ppxf, which were chosen randomly for every Monte Carlo realization.

I emphasize the fact that for the initial LOSVD convolution I did {\em not} use the analytic Fourier transform introduced in this paper, nor the analytic pixel integration of the gas emission lines. Instead I performed these steps numerically on the finely-sampled spectrum. This provides a useful debugging of my software implementation, for both gas and stars. In fact a good recovery can only be achieved if the analytic Fourier approach correctly and accurately corresponds to the numeric pixel integration and LOSVD convolution, in the limit of a well-sampled kernel, and when the spectrum contains information on the Gauss-Hermite coefficients, namely when  $\sigma\ga\sigma_{\rm inst}$.

\begin{figure*}
    \includegraphics[width=0.33\textwidth]{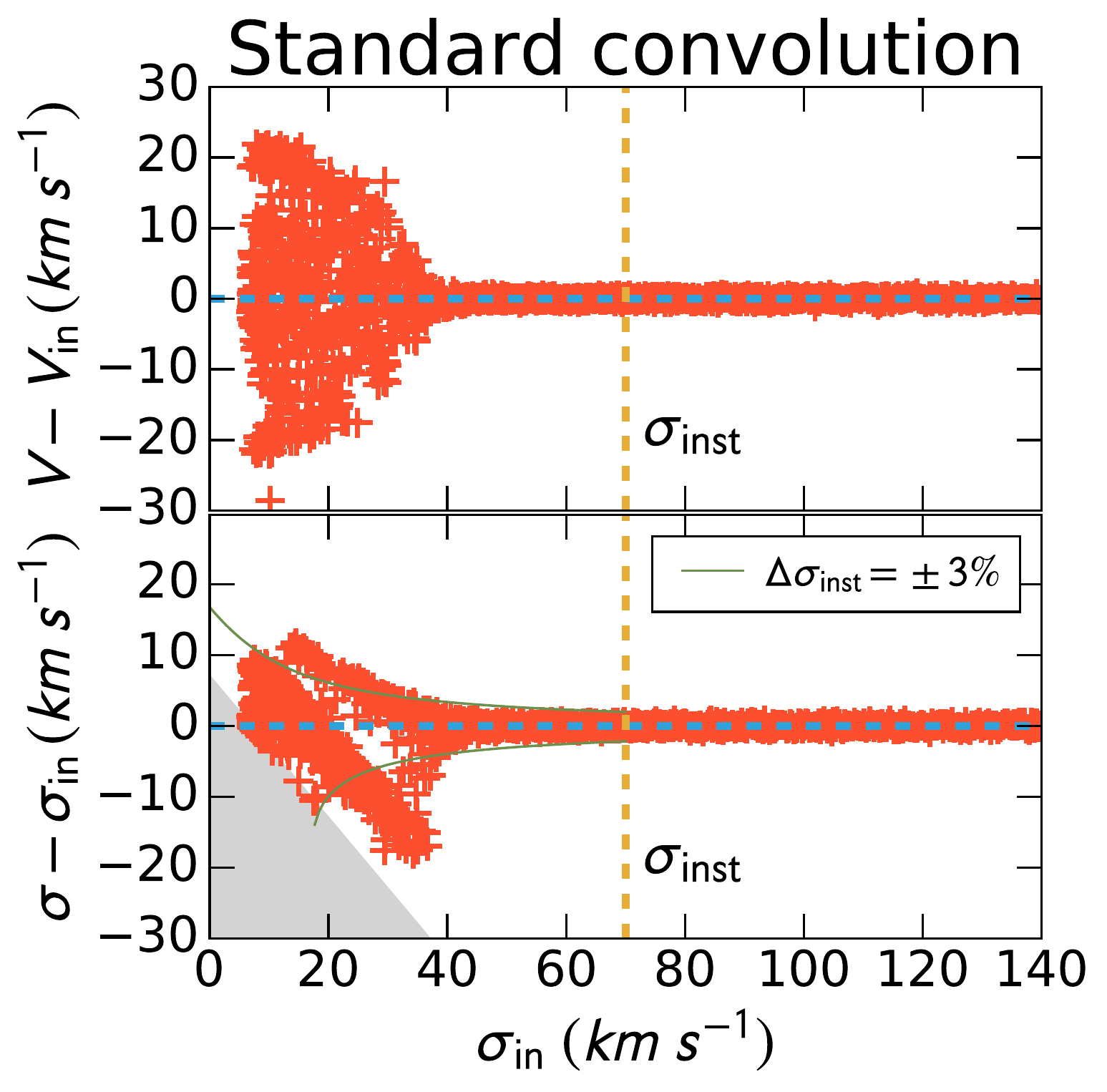}
    \includegraphics[width=0.33\textwidth]{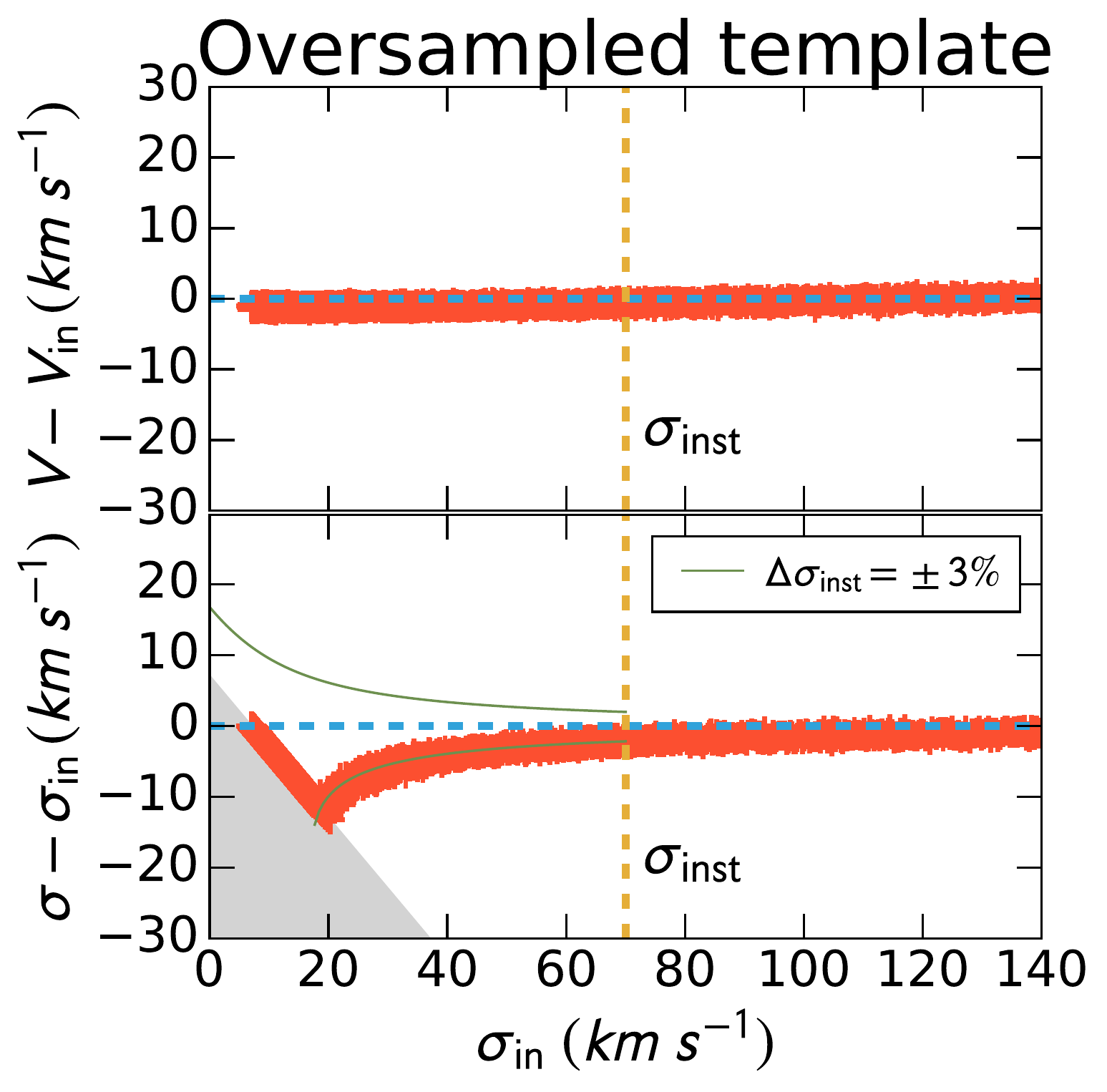}
    \includegraphics[width=0.33\textwidth]{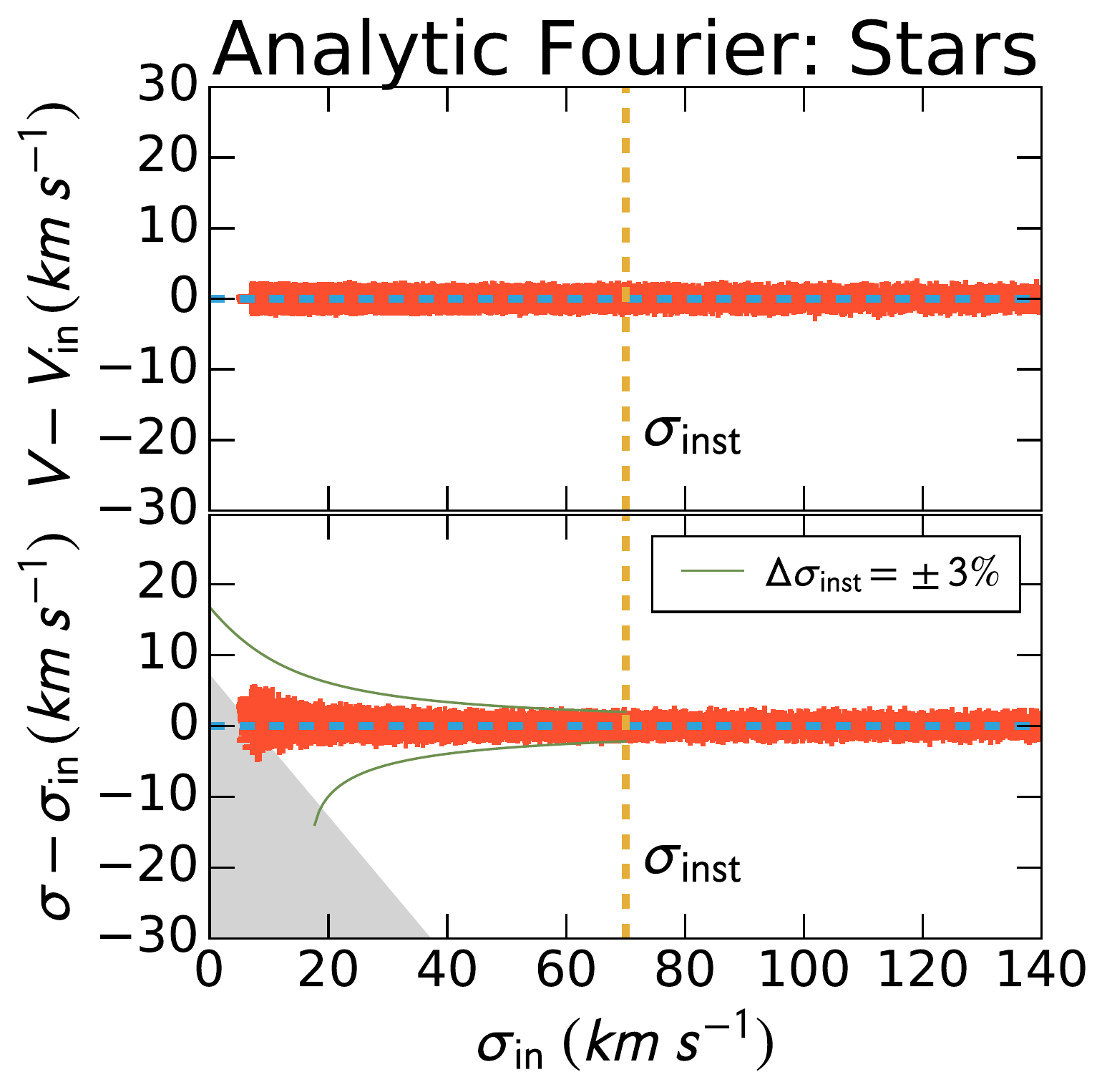}
    \caption{Stellar kinematics recovery with a Gaussian LOSVD. The three panels illustrate the \ppxf\ recovery of the velocity $V$ and dispersion $\sigma$ for an adopted Gaussian LOSVD with known velocity $V_{\rm in}$ and dispersion $\sigma_{\rm in}$. The vertical dashed line indicates the instrumental dispersion $\sigma_{\rm inst}=70$ \kms, which coincides with the adopted velocity sampling $(\Delta x)_{\rm pix}$. The grey region with $\sigma<\sigma_{\rm inst}/10$ is not allowed by the program.  The thin lines in the bottom panels illustrate the effect on $\sigma$ of an error in the instrumental dispersion $\Delta\sigma_{\rm inst}=\pm3\%$.
    {\em Left Panel:} recovery without oversampling, with the old \ppxf. Note the dramatic increase of the errors below $\sigma_{\rm in}\la\sigma_{\rm inst}/2$.
    {\em Middle Panel:} recovery with a well-sampled LOSVD and a template oversampled by a factor of 10, with the old \ppxf. This approach is slower and produces a significant drop in the dispersion below $\sigma_{\rm in}\la\sigma_{\rm inst}/2$.
    {\em Right Panel:} recovery with the solution proposed in this paper, namely using an analytic Fourier transform of the LOSVD, with the new \ppxf. This is at least as fast as the left panel, but here both $V$ and $\sigma$ are recovered without bias.
    \label{fig:ppxf_vs}}
\end{figure*}

\subsection{Description of the problem}
\label{sec:problem}

The convolution of the templates with the LOSVDs in \autoref{eq:model} was until now performed by \ppxf\ according to the standard discrete definition
\begin{equation}\label{eq:discrete_conv}
(T\convolution \mathcal{L})_p \equiv \sum_{q=-Q/2}^{Q/2} T_{q+p}\, \mathcal{L}_q,
\end{equation}
where $Q$ is the number of elements where the kernel $\mathcal{L}_q\equiv\mathcal{L}(c x_q)$ is non-zero and $T_q\equiv T(x_q)$. In practice, for computational efficiency, this convolution is performed using the standard, mathematically equivalent, Fourier approach \citep[e.g.][\S 13.1]{Press2007}
\begin{equation}\label{eq:fourier_convolution}
T\convolution \mathcal{L} = \mathcal{F}^{-1}\left[\mathcal{F}(T)\cdot\mathcal{F}(\mathcal{L})\right]
\end{equation}
where $\mathcal{F}$ is the Discrete Fourier Transform (DFT) and $\mathcal{F}^{-1}$ is its inverse, which can be computed efficiently with a number of operations proportional to $P\log P$ using the classic Fast Fourier Transform (FFT) algorithm \citep{cooley1965fft}. For efficiency, \ppxf\ pre-computes the $\mathcal{F}(T)$ of all the templates, as they do not change during the $\chi^2$ minimization, and uses the specific FFT for real input, Numpy's \texttt{fft.rfft}, which is based on \textsc{fftpack} \citep{fftpack}, to further decrease the FFT computation time by a factor of two.

For the discrete convolution, the LOSVD, which acts as convolution kernel, was until now sampled by \ppxf\ at discrete intervals $(\Delta x)_{\rm pixel}=(\Delta V)_{\rm pixel}/c$ before the computation of $\mathcal{F}(\mathcal{L})$. This straightforward approach of discretely sampling the kernel, represents the standard practice, when one needs to perform convolutions of spectra or images with Gaussian kernels. It is used e.g.\ in the \texttt{GAUSS\_SMOOTH} function in IDL, in the \texttt{ndimage.filters.gaussian\_filter} function of the Python package Scipy \citep{Scipy2001}, or in the \texttt{convolution.Gaussian1DKernel} function of  the Python package \textsc{Astropy} \citep{Astropy2013}.

However, it is clear that, when $\sigma\la(\Delta x)_{\rm pixel}$, where $\sigma$ is the dispersion of the Gaussian, the kernel starts becoming severely under-sampled and cannot be expected to accurately represent the LOSVD any more. As an illustration, when the Gaussian is centred on a pixel and $\sigma=(\Delta x)_{\rm pixel}/2$, the kernel has essentially only three non-zero elements $\mathcal{L}=(0.14,1,0.14)$, fully dominated by the middle one. I note here that integrating the LOSVD over the pixels instead of sampling it at the pixel centre, as done for a Gaussian in \autoref{eq:pixel_integration}, is {\em not} a correct solution to the undersampling problem. Instead, it is equivalent to incorrectly performing an extra convolution of the templates with a box function of the size of a pixel.

One should not expect a reliable recovery of the LOSVD at these low dispersion, and in particular, both the velocity and the dispersion should not be expected to be determined to better than about half a pixel. This was the reason why, until now, \ppxf\ included the  \texttt{oversample} keyword to analytically calculate a well-sampled LOSVD before convolving it with a template oversampled by interpolation onto an equally-fine grid. This approach however suffers from two limitations: (i) it requires a significant increase in the size of the template spectra, resulting in an increase of the computation time and (ii) the oversampling of the templates is necessarily smooth below the observed scale, and cannot represent the dense forest of thin absorption lines in real stellar spectra. In fact the smooth oversampling has a similar effect as an extra convolution of true spectrum with a box function of the size of one spectral pixel.

The left panel of \autoref{fig:ppxf_vs} illustrates the dramatic problems which appear when $\sigma_{\rm in}\la(\Delta x)_{\rm pixel}/2$, while adopting a Gaussian input LOSVD: \ppxf\ becomes essentially unable to reliably recover both the velocity and the dispersion of the LOSVD. In fact, although the velocity has an rms error of only 0.3 \kms, when $\sigma_{\rm in}>(\Delta x)_{\rm pixel}$, the errors become as high as 20 \kms\ at low $\sigma_{\rm in}$. This problem is not specific to \ppxf\ but will affect any spectral fitting program that uses the standard  \autoref{eq:discrete_conv} to define convolution!

The middle panel of \autoref{fig:ppxf_vs} shows that, as expected, by computing a well-sampled LOSVD and convolving it with an oversampled template, the velocity can be well recovered, however the dispersion tends to be under-estimated at low $\sigma_{\rm in}$, because the oversampled template is smoother than the real spectrum. In the common situation adopted for my tests, where $(\Delta x)_{\rm pixel}\approx\sigma_{\rm inst}$, the oversampling has a similar effect as a 3\%  over-estimation of $\sigma_{\rm inst}$.

Similar problems can be seen when the LOSVD is assumed to be described by a significantly non-Gaussian shape, parametrized by \autoref{eq:losvd}. I adopted as input some realistic values $h_3=h_4=0.1$ \citep[e.g.][]{Emsellem2004}. Note that, contrary to what we did in \citet{Cappellari2004}, here I assume the input LOSVD to be precisely described by the adopted parametrization, because I want to test the accurate recovery of the {\em known} input Gauss-Hermite moments. 

\begin{figure*}
    \includegraphics[width=0.33\textwidth]{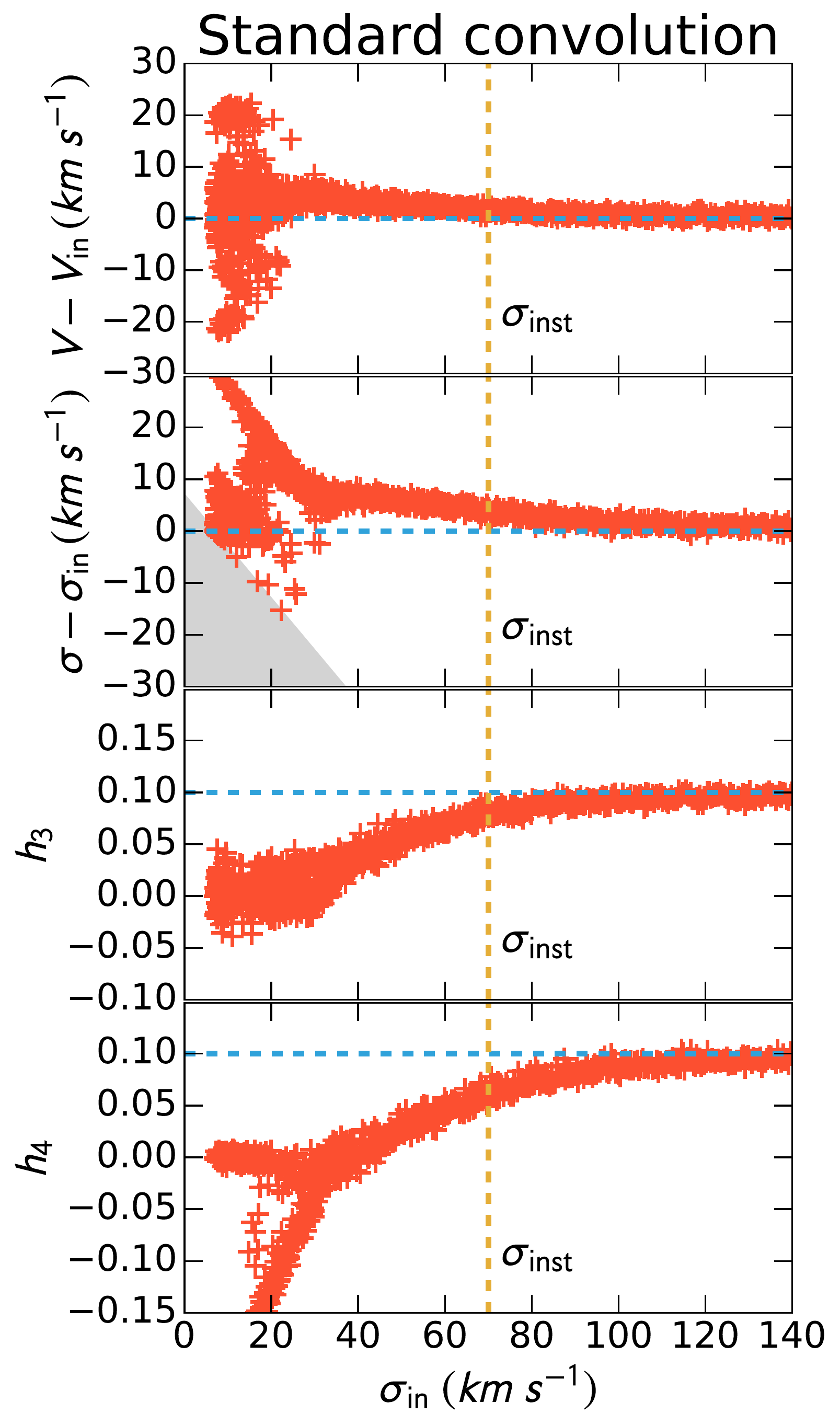}
    \includegraphics[width=0.33\textwidth]{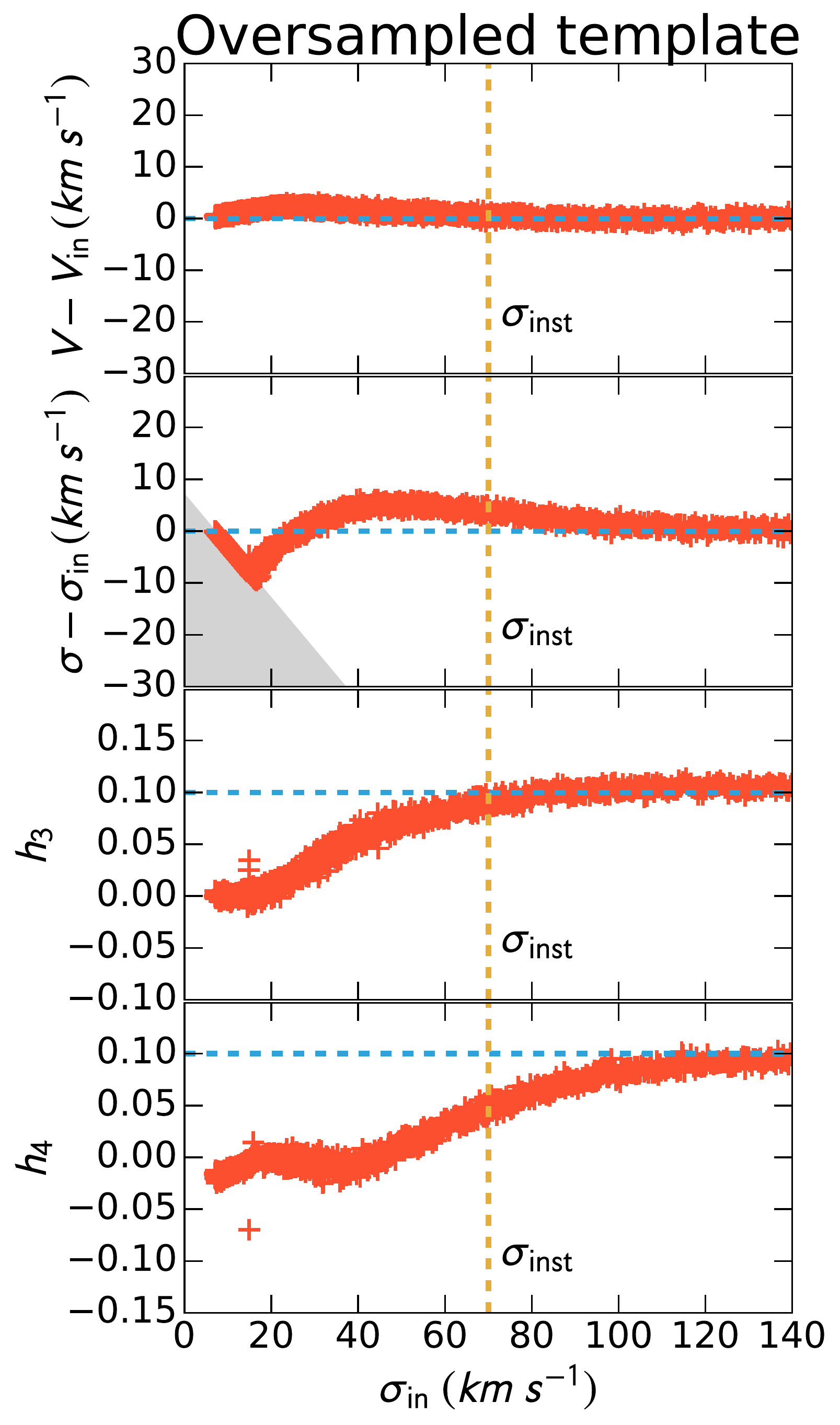}
    \includegraphics[width=0.33\textwidth]{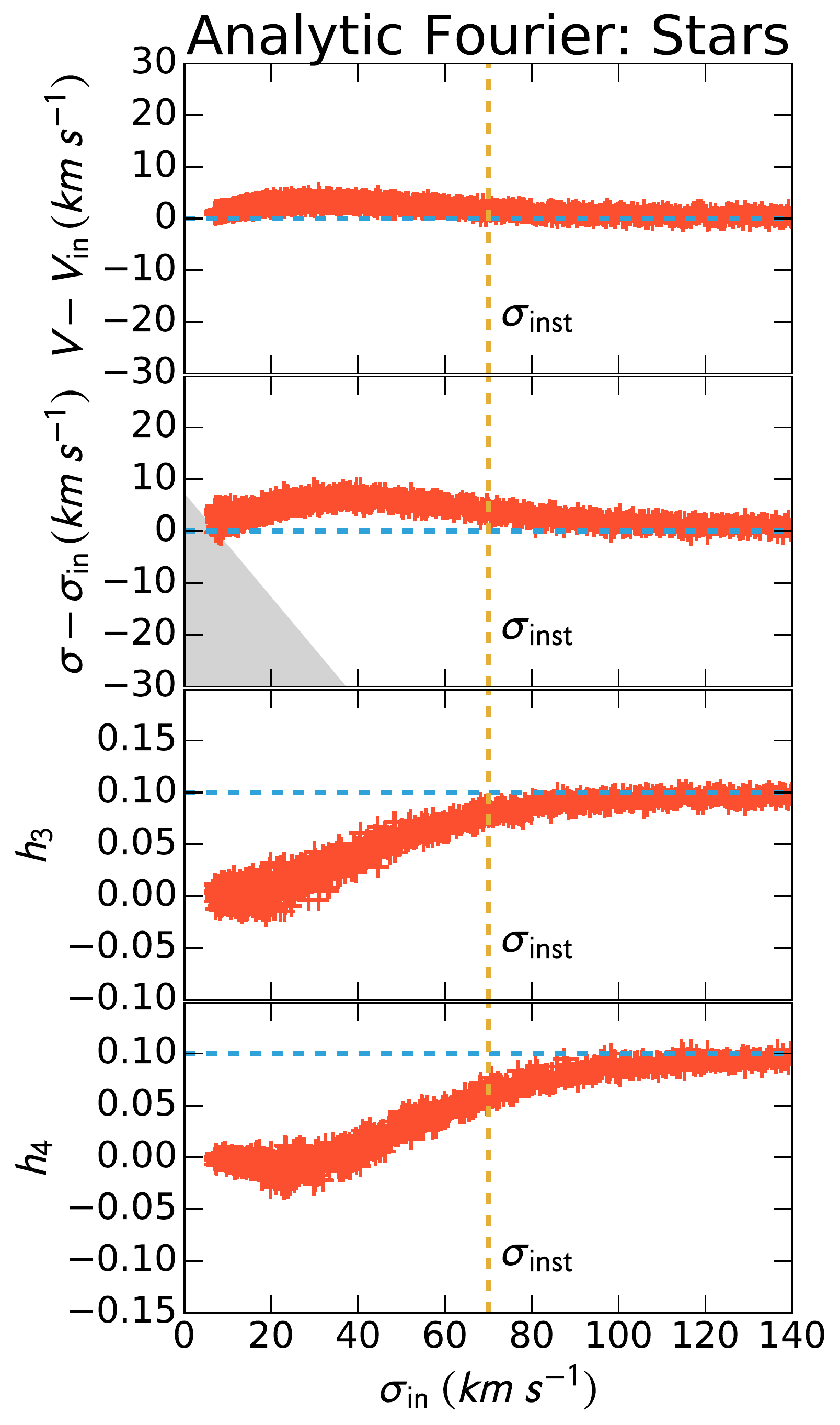}
    \caption{Stellar kinematics recovery with LOSVD described by the Gauss-Hermite parametrization. The three panels illustrate the \ppxf\ recovery of the kinematics for a LOSVD with adopted velocity $V_{\rm in}$, dispersion $\sigma_{\rm in}$ and realistic $h_3=h_4=0.1$. The vertical dashed line indicates the instrumental dispersion $\sigma_{\rm inst}=70$ \kms, which coincides with the adopted velocity sampling $(\Delta x)_{\rm pix}$. The grey region with $\sigma<\sigma_{\rm inst}/10$ is not allowed by the program. 
        {\em Left Panel:} recovery without oversampling, with the old \ppxf. Note the dramatic increase of the errors below $\sigma_{\rm in}\la\sigma_{\rm inst}/2$.
        {\em Middle Panel:} recovery with a well-sampled LOSVD and a template oversampled by a factor of 10, with the old \ppxf. This approach is slower and produces a significant drop in the dispersion below $\sigma_{\rm in}\la\sigma_{\rm inst}/2$.
        {\em Right Panel:} recovery with the solution proposed in this paper, namely using an analytic Fourier transform of the LOSVD, with the new \ppxf. This is at least as fast as the left panel, but here all kinematic parameters are properly recovered. Note that $h_3$ and $h_4$ start converging towards zero (i.e.\ Gaussian LOSVD) when $\sigma_{\rm in}\la\sigma_{\rm inst}$. This is intentional and unavoidable. It is used to prevent a dramatic increase of the errors in $V$ and $\sigma$ when the parameters of the LOSVD become degenerate at low $\sigma_{\rm in}$. This effect is extensively discussed in \citet{Cappellari2004}.
        \label{fig:ppxf_gh}}    
\end{figure*}

The left panel of \autoref{fig:ppxf_gh} again illustrates the dramatic problems in the recovery of the velocity and dispersion, when $\sigma_{\rm in}\la(\Delta x)_{\rm pixel}/2$. In this case, with non-Gaussian LOSVD, one can see the expected convergence towards zero of the $h_3$ and $h_4$ parameters, due to the penalty term in \autoref{eq:penalty}. For all the examples in \autoref{fig:ppxf_gh} I adopted the same penalty \texttt{bias=1} in \ppxf. This becomes important for $\sigma_{\rm in}\la(\Delta x)_{\rm pixel}$, where the broadening by the instrumental dispersion makes the deviations from a Gaussian hard to measure and \ppxf\ tries to penalize the LOSVD towards a Gaussian. But at the smallest $\sigma_{\rm in}$ even the penalty looses it effectiveness, and one can see sharp variations, especially in the recovered $h_4$, instead of the desired convergence toward zero (i.e.\ Gaussian shape).

The middle panel of \autoref{fig:ppxf_gh} again shows that by computing a well-sampled LOSVD and oversampling the template one can overcome the main problems with the recovery of the velocity. However the velocity dispersion shows the same behaviour as in the middle panel of \autoref{fig:ppxf_vs}, with the recovered $\sigma$ being underestimated, and hitting the lower boundary, when $\sigma_{\rm in}\la20$ \kms.

\subsection{Solution of the problem}

\begin{figure*}
    \includegraphics[width=0.33\textwidth]{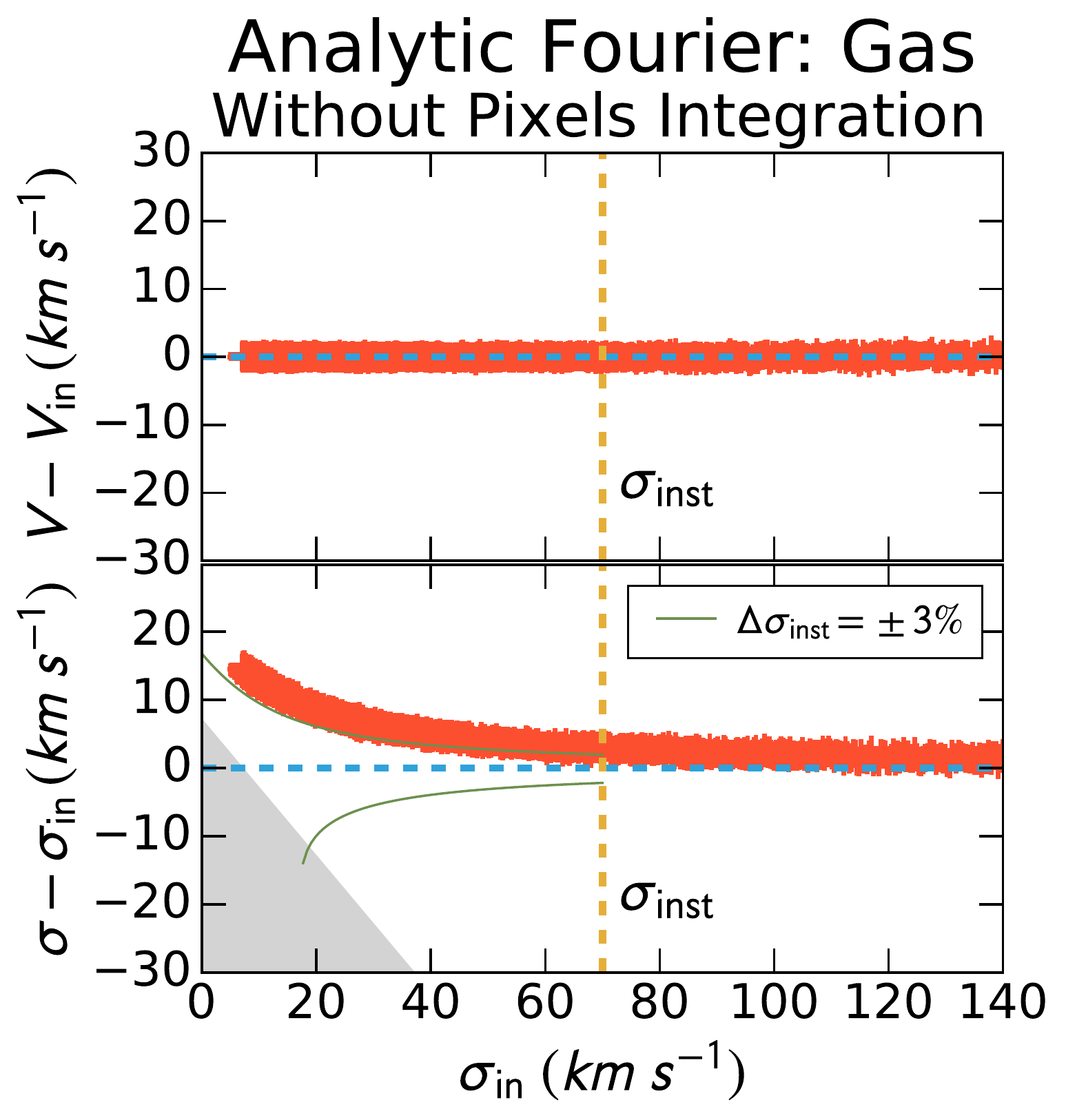}
    \includegraphics[width=0.33\textwidth]{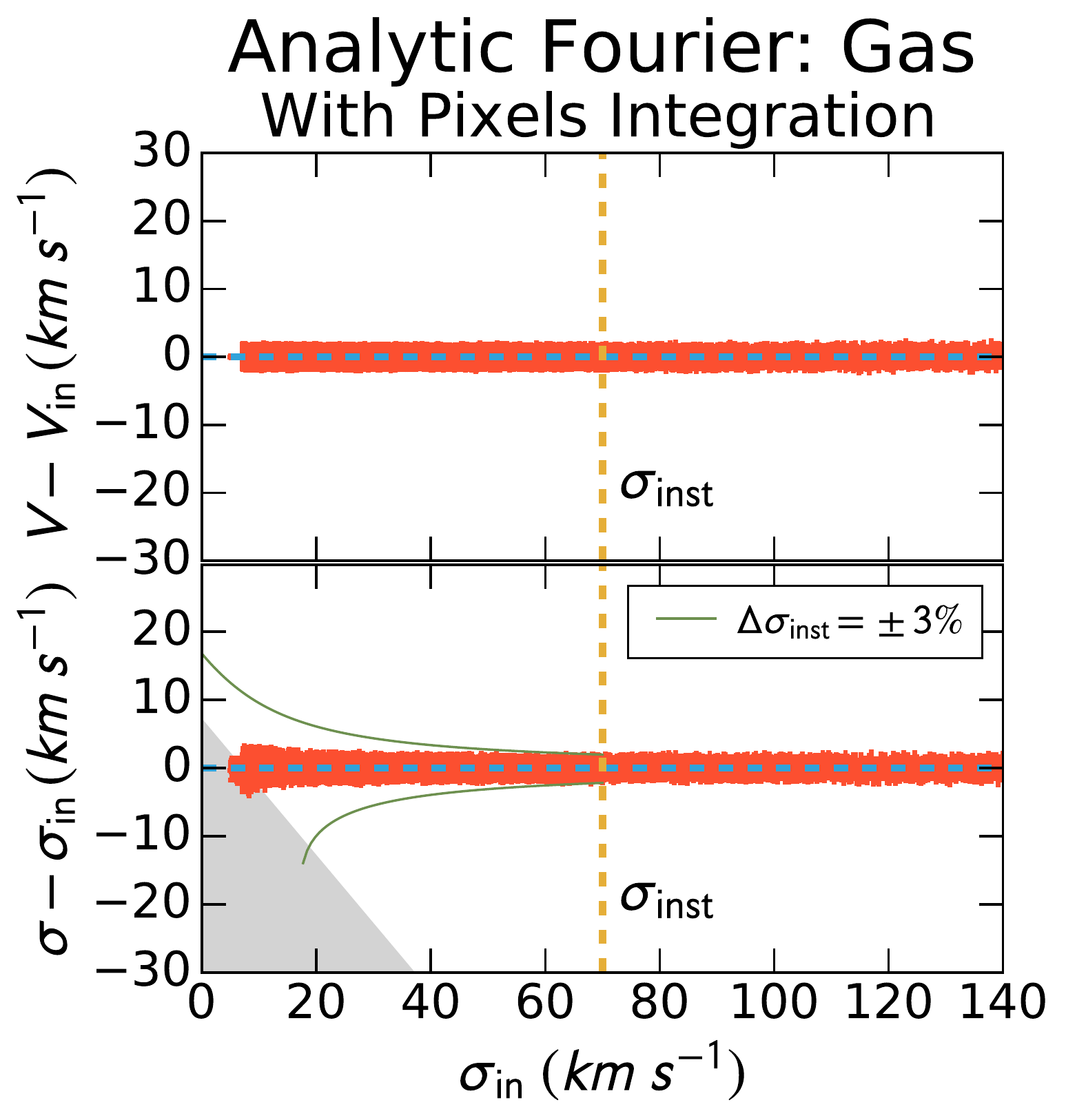}
    \caption{Gas kinematics with \ppxf. The two panels illustrate the recovery of the gas kinematics for the [\ion{O}{iii}]~$\lambda\lambda$4959,5007 doublet shown in \autoref{fig:solar_spectrum}. The input lines are assumed Gaussian, with adopted velocity $V_{\rm in}$ and dispersion $\sigma_{\rm in}$, and are fitted with Gaussian gas templates.
    The vertical dashed line indicates the instrumental dispersion $\sigma_{\rm inst}=70$ \kms, which coincides with the adopted velocity sampling $(\Delta x)_{\rm pix}$. The thin lines in the bottom panels illustrate the effect on $\sigma$ of an error in the instrumental dispersion $\Delta\sigma_{\rm inst}=\pm3\%$.
    {\em Left Panel:} recovery of gas kinematics while ignoring the integration of the Gaussian line over the spectral pixels. This is nearly equivalent to an underestimation of the $\sigma_{\rm inst}$.
    {\em Right Panel:} gas recovery with the proper integration of the Gaussian line over the pixels of \autoref{eq:pixel_integration}. The gas kinematics is now accurately recovered down to the smallest dispersion. 
    \label{fig:ppxf_gas}}    
\end{figure*}

A clean solution to the under-sampling problem comes by noting that the Fourier transform of the kernel happens to be analytic in the special case of interest, where the LOSVD is parametrized by the Gauss-Hermite series of \autoref{eq:losvd}. This suggests one can use an analytic rather than discrete Fourier transform to evaluate $\mathcal{F}(\mathcal{L})$ in \autoref{eq:fourier_convolution}.

If I define the continuous Fourier transform of a time-domain function $h(t)$, using the modern-physics convention, with $\omega$ the angular frequency, as
\begin{equation}
H(\omega)=\frac{1}{\sqrt{2\pi}}\int_{-\infty}^{\infty}h(t)\,e^{\mathrm{i}\omega t}\dd t,
\end{equation}
the $H(\omega)$ of the LOSVD has an exceedingly simple form. This is because the individual terms of $\mathcal{L}$ are eigenfunctions of the $H(\omega)$ operator. This fact was already pointed out by \citet[eq.~A7]{vanDerMarel93}, who quote the result from \citet[eq.~7.376.1]{Gradshteyn1980table}, and of course it can be verified analytically. It is a textbook result \citep[e.g.][\S 4]{schiff1968quantum}, related to the fact that the terms of \autoref{eq:losvd} are quantum-mechanical wave functions for the harmonic oscillator. These have the same form in either position or momentum-space, which are related via the Fourier transform.  Specifically, if one defines 
\begin{equation}
\mathcal{L}_m(t)=\frac{\exp\left(-t^2/2\right)}{\sigma\sqrt{2\pi}} H_m(t)
\end{equation}
the following simple transform pair holds
\begin{equation}\label{eq:ft_pair}
h(t)=\mathcal{L}_m(t)\iff H(\omega)=\mathrm{i}^m\,\mathcal{L}_m(\omega).
\end{equation}
In other words, the $H(\omega)$ have the same form as the original $\mathcal{L}_m$ functions, simply transformed into the frequency domain, while the even/odd terms are alternating real/imaginary respectively. This latter fact comes from two general properties of $H(\omega)$: (i) if $h(t)$ is real and even, then the same is true for $H(\omega)$; (ii) if $h(t)$ is real and odd, then $H(\omega)$ is imaginary and odd.

The analytic calculation of $\mathcal{F}(\mathcal{L})$ has two major advantages, with respect to its discrete version. They derive from general properties of the Fourier Transform  \citep[e.g.][\S 3]{brigham1974fft}:

\begin{description}
    \item {\bf Good kernel sampling:} When the LOSVD is narrow, ad consequently poorly sampled, in the `time' or spatial domain, it will become broad in the frequency domain according to the transform pair
    \begin{equation}\label{eq:freq_scaling}
    \frac{1}{|\sigma|}\,h\left(\frac{t}{\sigma}\right)\iff H(\sigma\omega)\qquad {\rm frequency~scaling}.
    \end{equation}
    This implies that the LOSVD will never be under-sampled in the frequency domain because the LOSVD is necessarily non-zero only for a small fraction of the size of the full template spectrum.
    
    \item {\bf Simple velocity shifting:} When evaluating the kernel of \autoref{eq:losvd} for the discrete convolution of \autoref{eq:discrete_conv}, one needed to sample a symmetric range of velocities from $-v_{\rm max}$ to $v_{\rm max}$, fully enclosing the offset LOSVD. This produced many zero elements in the kernel, causing practical issues with large $V$. In Fourier space this problem disappears as one can use
    \begin{equation}\label{eq:time_shifting}
    h(t-t_0)\iff H(\omega)\, e^{\mathrm{i}\omega t_0}\qquad {\rm time~shifting}
    \end{equation}
    to shift the velocity of the templates by simply changing the phase of all elements of the complex transform.        
\end{description}

\subsection{Final formula and tests}
\label{sec:formula}

Combining \autoref{eq:ft_pair} to \autoref{eq:time_shifting}, I find that the analytic Fourier transform of the LOSVD of \autoref{eq:losvd} is
\begin{equation}\label{eq:losvd_ft}
H(\omega) =
\frac{\exp\left(\mathrm{i}\,\omega V - \sigma^2\omega^2/2\right)}{\sqrt{2\pi}}
\left[1 + \sum_{m=3}^M \mathrm{i}^m h_m H_m(\sigma\omega) \right].
\end{equation}
This expression is used to compute the DFT $\mathcal{F}(\mathcal{L})$ which appears in \autoref{eq:fourier_convolution}.
Given that $\mathcal{L}$ is a real function, it is uniquely defined over half of the angular frequencies
\begin{equation}
h(t)\, {\rm is~real}\implies H(-\omega)=[H(\omega)]^\ast.
\end{equation}    
This implies that one only needs to evaluate $H(\omega)$ of \autoref{eq:losvd_ft} in the interval $\omega=[0,\pi]$ to obtain the DFT of real input. This needs to have the same number of elements as the DFT of real input $\mathcal{F}(T)$, namely half the size of $T$.

The right panel of \autoref{fig:ppxf_vs} illustrates the dramatic improvement in the recovery of both the velocity and the dispersion, when using the analytic Fourier transform approach. No biases can be detected in the recovery, and importantly, the scatter in the residuals is nearly independent of $\sigma_{\rm in}$. In particular, the velocity recovery is as accurate as in the case of the oversampled templates, down to the smallest $\sigma_{\rm in}$. Additionally, even the very small systematic trend in velocity, visible in the middle panel of \autoref{fig:ppxf_vs}, now completely disappears. The $\sigma$ is also recovered without detectable bias, and in particular, the underestimation of $\sigma$ that affects the oversampled case, also disappears.

Similar conclusions apply to the case with non-Gaussian LOSVD in the right panel of \autoref{fig:ppxf_gh}. Also in this case, \ppxf\ using the analytic $\mathcal{F}(\mathcal{L})$ looks similar to the oversampled case, but now the $\sigma$ underestimation disappears. Note that the penalty of \ppxf\ is still important to prevent wildly degenerate solutions and consequently large noise in all parameters when $\sigma_{\rm in}\la\sigma_{\rm inst}$. This important aspect was described in detail in \citet{Cappellari2004}. It is still relevant here, but the discussion will not be repeated.

In \autoref{fig:ppxf_gas} I illustrate the accuracy in the recovery of the gas kinematics using \ppxf\ with the analytic Fourier transform, and in particular here for the case of the [\ion{O}{iii}]~$\lambda\lambda$4959,5007 doublet. The results are the same for the other emission lines. The left panel shows the effect of neglecting the pixel integration of the Gaussian lines. In this case the bias in $\sigma$ is almost equivalent to an under-estimation of $\sigma_{\rm inst}$ of over 3\%, due to the missing convolution by a box function of the size of one spectral pixel. As one may have expected, the effect on $\sigma$ is nearly the opposite as that of oversampling the galaxy spectrum, in the middle panel of \autoref{fig:ppxf_vs}. The right panel of  \autoref{fig:ppxf_gas} presents the \ppxf\ recovery when including the analytic pixel integral of \autoref{eq:pixel_integration}. The gas kinematics is unbiased and recovered as accurately as the stellar one, at any $\sigma$.

Also shown, with thin lines, in the bottom panels of \autoref{fig:ppxf_vs} and \autoref{fig:ppxf_gas} is the variation in the recovered $\sigma$ one should expect when the assumed instrumental dispersion is in error by $\Delta\sigma_{\rm inst}=\pm3\%$, which represents the best accuracy achievable today \citep[e.g.][\S~7.4]{Yan2016}. The curves illustrate the fact that, at $\sigma\la\sigma_{\rm inst}/2$, biases in $\sigma$ will be dominated by the uncertainty in the LSF, which becomes a fundamental barrier to the accurate recovery of $\sigma$ in this regime. For this reason, it is important to emphasize that the key advantage of the proposed \ppxf\ upgrade is {\em not} the unbiased $\sigma$, but rather the fact that one can measure accurate velocities at any $\sigma$. The velocity is empirically an accurately defined quantity regardless of $\sigma$ and actually becomes even better defined at low $\sigma$. It is crucial for a kinematic extraction method to be able to keep providing reliable velocities even when $\sigma$ is too low to be measurable.

A complementary way of dealing with the undersampling of the LOSVD consists of using stellar templates observed with higher resolution that the galaxy, while keeping them at their native resolution. For this reason \ppxf\ was modified to allow for templates with a smaller velocity scale than the galaxy. It is also possible to skip the broadening of the templates to the galaxy LSF described in \autoref{sec:templates_library} and instead include the Gaussian convolution in \autoref{eq:losvd_ft} for $H(w)$, using the associative property of the convolution. An additional convolution by a Gaussian is obtained by multiplying $H(w)$ by the Gaussian Fourier transform. This implies that a convolution with the Gaussian of \autoref{eq:gauss_diff} is obtained by replacing the term inside the exponential in \autoref{eq:losvd_ft} with
\begin{equation}
\mathrm{i}\,\omega V - \left(\sigma_{\rm diff}^2+\sigma^2\right)\omega^2/2.
\end{equation}
To use this expression $\sigma_{\rm diff}$ must be constant in \kms. This must be achieved by homogenizing the template resolution in such a way that it has a constant $\sigma_{\rm diff}$ at all wavelengths. 

\subsection{Relation to previous work}

I have not found previous usage in the literature of an analytic Fourier transform to improve the accuracy of the spectral fitting method. However the general idea is obviously a good one and for this reason it was used before in the literature, in a similar context, for the special case of Gaussian functions.

The first generation of methods to extract the galaxies stellar kinematics generally performed the $\chi^2$ minimization in Fourier space, rather than in pixels space as \ppxf\ does. Given that the data to fit consisted of a Fourier transform, and it is common practice to use analytic fitting functions, it was natural to use as fitting function an analytic Fourier transform of the LOSVD, when available. In fact, this general idea, specifically the use of an analytic Fourier transform of a Gaussian broadening function, was employed by \citet{Sargent1977} and \citet{Schechter1979} for their improved version of the Fourier quotient method. The analytic transform was fitted in Fourier space to the ratio of the FFT of the galaxy and template. An analytic Fourier transform of the Gaussian was also used by \citet{Tonry1979} in their analysis of the errors of the cross-correlation method.

I found applications of an analytic Fourier transform for convolution in image processing. For example, the algorithm I describe in \autoref{sec:formula} for the Gauss-Hermite functions is a generalization of the algorithm~5 in the survey of Gaussian convolution algorithms by \cite{getreuer2013gaussian}. A key difference in this paper is that the templates we want to convolve are bandwidth limited (by the Gaussian LSF) and Nyquist sampled (at steps $\sigma_{\rm inst}$). Due to the sampling theorem, this implies the templates are {\em completely determined} by their samples. This is the reason we obtain an unbiased recovery of $\sigma$ down to the smallest values in the right panels of \autoref{fig:ppxf_vs} and \autoref{fig:ppxf_gas}, and we are not affected by the convolution bias at small $\sigma$ described by \citet{getreuer2013gaussian}, for generic non band-limited functions. 

In another broadly related work in image processing, \citet{Berger1999} noted the fact that two-dimensional Gauss-Hermite functions can be convolved analytically with Gaussians, thanks to their analytic Fourier transform, and propose to use this fact to accurately deconvolve images, to remove PSF effects.

\section{Summary}
\label{sec:summary}

In the first part of the paper, I provided an overview, or tutorial, of general concepts useful to understand and properly interpret the extraction of kinematics from galaxy spectra. I tried to clarify in particular the questions I received more often, over more than a decade, from users of my publicly available \ppxf\ software.

Then I gave an updated overview of the \ppxf\ method. I concentrated especially on the description of features of the method which were included after the publication of the original paper in 2004, some of which had never been properly explained and precisely documented in the literature.

Subsequently, I focused on the problem of extracting kinematics via full spectrum fitting, when the velocity dispersion is smaller than the spectral sampling, which is generally chosen to be the same as the instrumental dispersion. I illustrated the obvious but dramatic problems that arise when one completely ignores the issue, as well as the limitations of the previous solution, which consist of oversampling the spectra.

Finally I provided a clean solution to the long-standing under-sampling issue, which consist of using the analytic Fourier transform of the LOSVD in conjunction with the convolution theorem. This approach completely removes the need for oversampling and makes the full spectrum fitting method suitable for measuring reliable kinematics at any velocity dispersion. This is especially crucial for the mean velocity, which now becomes a well-determined quantity even when the dispersion becomes negligible, and consequently impossible to reliably recover from real data. 

The approach described in this paper was implemented in a significant upgrade to the publicly available \ppxf\ code, and is already being used as part of the MaNGA Data Analysis Pipeline (Wesfall et al.\ in preparation). Further tests on real IFS data will be published elsewhere.

The proposed solution appears quite natural, however, perhaps surprisingly, it is currently not being used by any of the popular software packages. Given the simplicity of our approach for accurate convolutions, we argue it should become standard practice.

\section*{Acknowledgements}

This paper was motivated by productive discussions within the MaNGA Data Analysis Pipeline team and in particular with Kyle Wesfall and Matt Bershady. I thank Samantha Penny for starting the discussion on this subject. I am also grateful to Eric Emsellem and Daniel Thomas for comments. I acknowledge support from a Royal Society University Research Fellowship. This paper made use of \textsc{Matplotlib} \citep{matplotlib2007} and of the \textsc{lineid\_plot}\footnote{Available from \url{https://github.com/phn/lineid_plot}}  Python program by Prasanth Nair.

\bibliographystyle{mnras}
\bibliography{cappellari_ppxf_fourier}

\begin{thebibliography}{}
\makeatletter
\relax
\def\mn@urlcharsother{\let\do\@makeother \do\$\do\&\do\#\do\^\do\_\do\%\do\~}
\def\mn@doi{\begingroup\mn@urlcharsother \@ifnextchar [ {\mn@doi@}
  {\mn@doi@[]}}
\def\mn@doi@[#1]#2{\def\@tempa{#1}\ifx\@tempa\@empty \href
  {http://dx.doi.org/#2} {doi:#2}\else \href {http://dx.doi.org/#2} {#1}\fi
  \endgroup}
\def\mn@eprint#1#2{\mn@eprint@#1:#2::\@nil}
\def\mn@eprint@arXiv#1{\href {http://arxiv.org/abs/#1} {{\tt arXiv:#1}}}
\def\mn@eprint@dblp#1{\href {http://dblp.uni-trier.de/rec/bibtex/#1.xml}
  {dblp:#1}}
\def\mn@eprint@#1:#2:#3:#4\@nil{\def\@tempa {#1}\def\@tempb {#2}\def\@tempc
  {#3}\ifx \@tempc \@empty \let \@tempc \@tempb \let \@tempb \@tempa \fi \ifx
  \@tempb \@empty \def\@tempb {arXiv}\fi \@ifundefined
  {mn@eprint@\@tempb}{\@tempb:\@tempc}{\expandafter \expandafter \csname
  mn@eprint@\@tempb\endcsname \expandafter{\@tempc}}}

\bibitem[\protect\citeauthoryear{{Abramowitz} \& {Stegun}}{{Abramowitz} \&
  {Stegun}}{1964}]{Abramowitz1964}
{Abramowitz} M.,  {Stegun} I.~A.,  1964, Handbook of Mathematical Functions
  (Reprinted 1972).
National Bureau of Standards, Washington, \url
  {https://books.google.com/books?id=MtU8uP7XMvoC}

\bibitem[\protect\citeauthoryear{{Alatalo} et~al.,}{{Alatalo}
  et~al.}{2011}]{Alatalo2011}
{Alatalo} K.,  et~al., 2011, \mn@doi [\apj] {10.1088/0004-637X/735/2/88}, \href
  {http://adsabs.harvard.edu/abs/2011ApJ...735...88A} {735, 88}

\bibitem[\protect\citeauthoryear{{Astropy Collaboration}}{{Astropy
  Collaboration}}{2013}]{Astropy2013}
{Astropy Collaboration} 2013, \mn@doi [\aap] {10.1051/0004-6361/201322068},
  \href {http://adsabs.harvard.edu/abs/2013A%26A...558A..33A} {558, A33}

\bibitem[\protect\citeauthoryear{{Barrera-Ballesteros}
  et~al.,}{{Barrera-Ballesteros} et~al.}{2015}]{Barrera-Ballesteros2015}
{Barrera-Ballesteros} J.~K.,  et~al., 2015, \mn@doi [\aap]
  {10.1051/0004-6361/201424935}, \href
  {http://adsabs.harvard.edu/abs/2015A%26A...582A..21B} {582, A21}

\bibitem[\protect\citeauthoryear{{Bender}, {Saglia}  \& {Gerhard}}{{Bender}
  et~al.}{1994}]{Bender1994}
{Bender} R.,  {Saglia} R.~P.,   {Gerhard} O.~E.,  1994, \mnras, \href
  {http://adsabs.harvard.edu/abs/1994MNRAS.269..785B} {269, 785}

\bibitem[\protect\citeauthoryear{{Berger} \& {Simental}}{{Berger} \&
  {Simental}}{1999}]{Berger1999}
{Berger} H.,  {Simental} E.,  1999, in {Descour} M.~R.,  {Shen} S.~S.,  eds,
  \procspie\ Vol. 3753, Imaging Spectrometry V. pp 124--132,
  \mn@doi{10.1117/12.366275}

\bibitem[\protect\citeauthoryear{{Bershady}, {Verheijen}, {Swaters},
  {Andersen}, {Westfall}  \& {Martinsson}}{{Bershady}
  et~al.}{2010}]{Bershady2010}
{Bershady} M.~A.,  {Verheijen} M.~A.~W.,  {Swaters} R.~A.,  {Andersen} D.~R.,
  {Westfall} K.~B.,   {Martinsson} T.,  2010, \mn@doi [\apj]
  {10.1088/0004-637X/716/1/198}, \href
  {http://adsabs.harvard.edu/abs/2010ApJ...716..198B} {716, 198}

\bibitem[\protect\citeauthoryear{{Blanc} et~al.,}{{Blanc}
  et~al.}{2013}]{Blanc2013}
{Blanc} G.~A.,  et~al., 2013, \mn@doi [\aj] {10.1088/0004-6256/145/5/138},
  \href {http://adsabs.harvard.edu/abs/2013AJ....145..138B} {145, 138}

\bibitem[\protect\citeauthoryear{Branch, Coleman  \& Li}{Branch
  et~al.}{1999}]{branch1999subspace}
Branch M.~A.,  Coleman T.~F.,   Li Y.,  1999, \mn@doi [SIAM Journal on
  Scientific Computing] {10.1137/S1064827595289108}, 21, 1

\bibitem[\protect\citeauthoryear{Brigham}{Brigham}{1974}]{brigham1974fft}
Brigham E.~O.,  1974, The fast Fourier transform.
Prentice-Hall Inc., Englewood Cliffs, NJ

\bibitem[\protect\citeauthoryear{{Bruzual} \& {Charlot}}{{Bruzual} \&
  {Charlot}}{2003}]{bruzual03}
{Bruzual} G.,  {Charlot} S.,  2003, \mn@doi [\mnras]
  {10.1046/j.1365-8711.2003.06897.x}, \href
  {http://adsabs.harvard.edu/abs/2003MNRAS.344.1000B} {344, 1000}

\bibitem[\protect\citeauthoryear{{Bryant} et~al.,}{{Bryant}
  et~al.}{2015}]{Bryant2015}
{Bryant} J.~J.,  et~al., 2015, \mn@doi [\mnras] {10.1093/mnras/stu2635}, \href
  {http://adsabs.harvard.edu/abs/2015MNRAS.447.2857B} {447, 2857}

\bibitem[\protect\citeauthoryear{{Bundy} et~al.,}{{Bundy}
  et~al.}{2015}]{Bundy2015}
{Bundy} K.,  et~al., 2015, \mn@doi [\apj] {10.1088/0004-637X/798/1/7}, \href
  {http://adsabs.harvard.edu/abs/2015ApJ...798....7B} {798, 7}

\bibitem[\protect\citeauthoryear{{Calzetti}, {Armus}, {Bohlin}, {Kinney},
  {Koornneef}  \& {Storchi-Bergmann}}{{Calzetti} et~al.}{2000}]{Calzetti2000}
{Calzetti} D.,  {Armus} L.,  {Bohlin} R.~C.,  {Kinney} A.~L.,  {Koornneef} J.,
   {Storchi-Bergmann} T.,  2000, \mn@doi [\apj] {10.1086/308692}, \href
  {http://adsabs.harvard.edu/abs/2000ApJ...533..682C} {533, 682}

\bibitem[\protect\citeauthoryear{{Cappellari}}{{Cappellari}}{2016}]{Cappellari2016}
{Cappellari} M.,  2016, \mn@doi [\araa] {10.1146/annurev-astro-082214-122432},
  \href {http://adsabs.harvard.edu/abs/2016ARA%26A..54..597C} {54, 597}

\bibitem[\protect\citeauthoryear{{Cappellari} \& {Emsellem}}{{Cappellari} \&
  {Emsellem}}{2004}]{Cappellari2004}
{Cappellari} M.,  {Emsellem} E.,  2004, \mn@doi [\pasp] {10.1086/381875}, \href
  {http://adsabs.harvard.edu/abs/2004PASP..116..138C} {116, 138}

\bibitem[\protect\citeauthoryear{{Cappellari} et~al.,}{{Cappellari}
  et~al.}{2009}]{Cappellari2009}
{Cappellari} M.,  et~al., 2009, \mn@doi [\apjl] {10.1088/0004-637X/704/1/L34},
  \href {http://adsabs.harvard.edu/abs/2009ApJ...704L..34C} {704, L34}

\bibitem[\protect\citeauthoryear{{Cappellari} et~al.,}{{Cappellari}
  et~al.}{2011}]{Cappellari2011a}
{Cappellari} M.,  et~al., 2011, \mn@doi [\mnras]
  {10.1111/j.1365-2966.2010.18174.x}, \href
  {http://adsabs.harvard.edu/abs/2011MNRAS.413..813C} {413, 813}

\bibitem[\protect\citeauthoryear{{Cappellari} et~al.,}{{Cappellari}
  et~al.}{2012}]{Cappellari2012}
{Cappellari} M.,  et~al., 2012, \mn@doi [\nat] {10.1038/nature10972}, \href
  {http://adsabs.harvard.edu/abs/2012Natur.484..485C} {484, 485}

\bibitem[\protect\citeauthoryear{{Cappellari} et~al.,}{{Cappellari}
  et~al.}{2013a}]{Cappellari2013p15}
{Cappellari} M.,  et~al., 2013a, \mn@doi [\mnras] {10.1093/mnras/stt562}, \href
  {http://adsabs.harvard.edu/abs/2013MNRAS.432.1709C} {432, 1709}

\bibitem[\protect\citeauthoryear{{Cappellari} et~al.,}{{Cappellari}
  et~al.}{2013b}]{Cappellari2013p20}
{Cappellari} M.,  et~al., 2013b, \mn@doi [\mnras] {10.1093/mnras/stt644}, \href
  {http://adsabs.harvard.edu/abs/2013MNRAS.432.1862C} {432, 1862}

\bibitem[\protect\citeauthoryear{{Cappellari} et~al.,}{{Cappellari}
  et~al.}{2015}]{Cappellari2015dm}
{Cappellari} M.,  et~al., 2015, \mn@doi [\apjl] {10.1088/2041-8205/804/1/L21},
  \href {http://adsabs.harvard.edu/abs/2015ApJ...804L..21C} {804, L21}

\bibitem[\protect\citeauthoryear{{Cardelli}, {Clayton}  \& {Mathis}}{{Cardelli}
  et~al.}{1989}]{Cardelli1989}
{Cardelli} J.~A.,  {Clayton} G.~C.,   {Mathis} J.~S.,  1989, \mn@doi [\apj]
  {10.1086/167900}, \href {http://adsabs.harvard.edu/abs/1989ApJ...345..245C}
  {345, 245}

\bibitem[\protect\citeauthoryear{{Cenarro}, {Cardiel}, {Gorgas}, {Peletier},
  {Vazdekis}  \& {Prada}}{{Cenarro} et~al.}{2001}]{Cenarro2001}
{Cenarro} A.~J.,  {Cardiel} N.,  {Gorgas} J.,  {Peletier} R.~F.,  {Vazdekis}
  A.,   {Prada} F.,  2001, \mn@doi [\mnras] {10.1046/j.1365-8711.2001.04688.x},
  \href {http://adsabs.harvard.edu/abs/2001MNRAS.326..959C} {326, 959}

\bibitem[\protect\citeauthoryear{{Cheung} et~al.,}{{Cheung}
  et~al.}{2016}]{Cheung2016}
{Cheung} E.,  et~al., 2016, \mn@doi [\nat] {10.1038/nature18006}, \href
  {http://adsabs.harvard.edu/abs/2016Natur.533..504C} {533, 504}

\bibitem[\protect\citeauthoryear{{Cid Fernandes}, {Mateus}, {Sodr{\'e}},
  {Stasi{\'n}ska}  \& {Gomes}}{{Cid Fernandes} et~al.}{2005}]{CidFernandes2005}
{Cid Fernandes} R.,  {Mateus} A.,  {Sodr{\'e}} L.,  {Stasi{\'n}ska} G.,
  {Gomes} J.~M.,  2005, \mn@doi [\mnras] {10.1111/j.1365-2966.2005.08752.x},
  \href {http://adsabs.harvard.edu/abs/2005MNRAS.358..363C} {358, 363}

\bibitem[\protect\citeauthoryear{{Coelho}, {Barbuy}, {Mel{\'e}ndez}, {Schiavon}
   \& {Castilho}}{{Coelho} et~al.}{2005}]{Coelho2005}
{Coelho} P.,  {Barbuy} B.,  {Mel{\'e}ndez} J.,  {Schiavon} R.~P.,   {Castilho}
  B.~V.,  2005, \mn@doi [\aap] {10.1051/0004-6361:20053511}, \href
  {http://adsabs.harvard.edu/abs/2005A%26A...443..735C} {443, 735}

\bibitem[\protect\citeauthoryear{{Conroy}}{{Conroy}}{2013}]{Conroy2013}
{Conroy} C.,  2013, \mn@doi [\araa] {10.1146/annurev-astro-082812-141017},
  \href {http://adsabs.harvard.edu/abs/2013ARA%26A..51..393C} {51, 393}

\bibitem[\protect\citeauthoryear{{Conroy} \& {van Dokkum}}{{Conroy} \& {van
  Dokkum}}{2012}]{Conroy2012models}
{Conroy} C.,  {van Dokkum} P.,  2012, \mn@doi [\apj]
  {10.1088/0004-637X/747/1/69}, \href
  {http://adsabs.harvard.edu/abs/2012ApJ...747...69C} {747, 69}

\bibitem[\protect\citeauthoryear{Cooley \& Tukey}{Cooley \&
  Tukey}{1965}]{cooley1965fft}
Cooley J.~W.,  Tukey J.~W.,  1965, \mn@doi [Mathematics of computation]
  {10.2307/2003354}, 19, 297

\bibitem[\protect\citeauthoryear{{Davis} et~al.,}{{Davis}
  et~al.}{2011}]{Davis2011b}
{Davis} T.~A.,  et~al., 2011, \mn@doi [\mnras]
  {10.1111/j.1365-2966.2011.19355.x}, \href
  {http://adsabs.harvard.edu/abs/2011MNRAS.417..882D} {417, 882}

\bibitem[\protect\citeauthoryear{{Emsellem} et~al.,}{{Emsellem}
  et~al.}{2004}]{Emsellem2004}
{Emsellem} E.,  et~al., 2004, \mn@doi [\mnras]
  {10.1111/j.1365-2966.2004.07948.x}, \href
  {http://adsabs.harvard.edu/abs/2004MNRAS.352..721E} {352, 721}

\bibitem[\protect\citeauthoryear{{Emsellem} et~al.,}{{Emsellem}
  et~al.}{2011}]{Emsellem2011}
{Emsellem} E.,  et~al., 2011, \mn@doi [\mnras]
  {10.1111/j.1365-2966.2011.18496.x}, \href
  {http://adsabs.harvard.edu/abs/2011MNRAS.414..888E} {414, 888}

\bibitem[\protect\citeauthoryear{Gelman, Carlin, Stern, Dunson, Vehtari  \&
  Rubin}{Gelman et~al.}{2013}]{gelman2013bayesian}
Gelman A.,  Carlin J.~B.,  Stern H.~S.,  Dunson D.~B.,  Vehtari A.,   Rubin
  D.~B.,  2013, Bayesian data analysis, 3rd edition.
Chapman \& Hall/CRC, Boca Raton, \url
  {https://books.google.com/books?id=ZXL6AQAAQBAJ}

\bibitem[\protect\citeauthoryear{{Gerhard}}{{Gerhard}}{1993}]{Gerhard1993}
{Gerhard} O.~E.,  1993, \mnras, \href
  {http://adsabs.harvard.edu/abs/1993MNRAS.265..213G} {265, 213}

\bibitem[\protect\citeauthoryear{Getreuer}{Getreuer}{2013}]{getreuer2013gaussian}
Getreuer P.,  2013, \mn@doi [Image Processing On Line] {10.5201/ipol.2013.87},
  3, 286

\bibitem[\protect\citeauthoryear{Gradshteyn \& Ryzhik}{Gradshteyn \&
  Ryzhik}{1980}]{Gradshteyn1980table}
Gradshteyn I.~S.,  Ryzhik I.~M.,  1980, Table of integrals, series, and
  products (8th ed.\ 2014).
Academic Press, New York, \mn@doi{10.1016/B978-0-12-384933-5.00008-4}

\bibitem[\protect\citeauthoryear{{Gustafsson}, {Edvardsson}, {Eriksson},
  {J{\o}rgensen}, {Nordlund}  \& {Plez}}{{Gustafsson}
  et~al.}{2008}]{Gustafsson2008}
{Gustafsson} B.,  {Edvardsson} B.,  {Eriksson} K.,  {J{\o}rgensen} U.~G.,
  {Nordlund} {\AA}.,   {Plez} B.,  2008, \mn@doi [\aap]
  {10.1051/0004-6361:200809724}, \href
  {http://adsabs.harvard.edu/abs/2008A%26A...486..951G} {486, 951}

\bibitem[\protect\citeauthoryear{Hansen}{Hansen}{1998}]{hansen1998regul}
Hansen P.~C.,  1998, Rank-deficient and discrete ill-posed problems: numerical
  aspects of linear inversion.
 Mathematical Modeling and Computation Vol. 4, Siam, Philadelphia,
  \mn@doi{10.1137/1.9780898719697}

\bibitem[\protect\citeauthoryear{{Ho}, {Filippenko}  \& {Sargent}}{{Ho}
  et~al.}{1997}]{Ho1997}
{Ho} L.~C.,  {Filippenko} A.~V.,   {Sargent} W.~L.~W.,  1997, \mn@doi [\apjs]
  {10.1086/313041}, \href {http://adsabs.harvard.edu/abs/1997ApJS..112..315H}
  {112, 315}

\bibitem[\protect\citeauthoryear{{Ho} et~al.,}{{Ho} et~al.}{2016}]{Ho2016}
{Ho} I.-T.,  et~al., 2016, \mn@doi [\mnras] {10.1093/mnras/stw017}, \href
  {http://adsabs.harvard.edu/abs/2016MNRAS.457.1257H} {457, 1257}

\bibitem[\protect\citeauthoryear{{Hogg}}{{Hogg}}{1999}]{Hogg1999}
{Hogg} D.~W.,  1999, preprint, \href
  {http://adsabs.harvard.edu/abs/1999astro.ph..5116H} {} (\mn@eprint {}
  {astro-ph/9905116})

\bibitem[\protect\citeauthoryear{Hunter}{Hunter}{2007}]{matplotlib2007}
Hunter J.~D.,  2007, \mn@doi [Computing In Science \& Engineering]
  {10.1109/MCSE.2007.55}, 9, 90

\bibitem[\protect\citeauthoryear{{Johnston}, {Merrifield},
  {Arag{\'o}n-Salamanca}  \& {Cappellari}}{{Johnston}
  et~al.}{2013}]{Johnston2013}
{Johnston} E.~J.,  {Merrifield} M.~R.,  {Arag{\'o}n-Salamanca} A.,
  {Cappellari} M.,  2013, \mn@doi [\mnras] {10.1093/mnras/sts121}, \href
  {http://adsabs.harvard.edu/abs/2013MNRAS.428.1296J} {428, 1296}

\bibitem[\protect\citeauthoryear{Jones, Oliphant, Peterson  et~al.}{Jones
  et~al.}{2001}]{Scipy2001}
Jones E.,  Oliphant T.,  Peterson P.,   et~al., 2001, {SciPy}: Open source
  scientific tools for {Python}, \url {http://www.scipy.org/}

\bibitem[\protect\citeauthoryear{{Kelson}, {Illingworth}, {van Dokkum}  \&
  {Franx}}{{Kelson} et~al.}{2000}]{Kelson2000}
{Kelson} D.~D.,  {Illingworth} G.~D.,  {van Dokkum} P.~G.,   {Franx} M.,  2000,
  \mn@doi [\apj] {10.1086/308445}, \href
  {http://adsabs.harvard.edu/abs/2000ApJ...531..159K} {531, 159}

\bibitem[\protect\citeauthoryear{{Krajnovi{\'c}}, {Cappellari}, {de Zeeuw}  \&
  {Copin}}{{Krajnovi{\'c}} et~al.}{2006}]{Krajnovic2006}
{Krajnovi{\'c}} D.,  {Cappellari} M.,  {de Zeeuw} P.~T.,   {Copin} Y.,  2006,
  \mn@doi [\mnras] {10.1111/j.1365-2966.2005.09902.x}, \href
  {http://adsabs.harvard.edu/abs/2006MNRAS.366..787K} {366, 787}

\bibitem[\protect\citeauthoryear{{Krajnovi{\'c}}, {McDermid}, {Cappellari}  \&
  {Davies}}{{Krajnovi{\'c}} et~al.}{2009}]{Krajnovic2009}
{Krajnovi{\'c}} D.,  {McDermid} R.~M.,  {Cappellari} M.,   {Davies} R.~L.,
  2009, \mn@doi [\mnras] {10.1111/j.1365-2966.2009.15415.x}, \href
  {http://adsabs.harvard.edu/abs/2009MNRAS.399.1839K} {399, 1839}

\bibitem[\protect\citeauthoryear{{Kurucz}}{{Kurucz}}{2005}]{Kurucz2005}
{Kurucz} R.~L.,  2005, Memorie della Societa Astronomica Italiana Supplementi,
  \href {http://adsabs.harvard.edu/abs/2005MSAIS...8..189K} {8, 189}

\bibitem[\protect\citeauthoryear{Lawson \& Hanson}{Lawson \&
  Hanson}{1974}]{Lawson1995}
Lawson C.~L.,  Hanson R.~J.,  1974, Solving least squares problems (SIAM 1995
  edition).
 Classics in applied mathematics Vol. 15, Prentice-Hall Inc., Englewood Cliffs,
  NJ, \mn@doi{10.1137/1.9781611971217}

\bibitem[\protect\citeauthoryear{{Maraston} \& {Str{\"o}mb{\"a}ck}}{{Maraston}
  \& {Str{\"o}mb{\"a}ck}}{2011}]{Maraston2011}
{Maraston} C.,  {Str{\"o}mb{\"a}ck} G.,  2011, \mn@doi [\mnras]
  {10.1111/j.1365-2966.2011.19738.x}, \href
  {http://adsabs.harvard.edu/abs/2011MNRAS.418.2785M} {418, 2785}

\bibitem[\protect\citeauthoryear{{Markwardt}}{{Markwardt}}{2009}]{Markwardt2009}
{Markwardt} C.~B.,  2009, in D.~A.~Bohlender D.~Durand .~P.~D.,  ed.,
  Astronomical Society of the Pacific Conference Series Vol. 411, Astronomical
  Data Analysis Software and Systems XVIII. p.~251 (\mn@eprint {arXiv}
  {0902.2850})

\bibitem[\protect\citeauthoryear{{McDermid} et~al.,}{{McDermid}
  et~al.}{2015}]{McDermid2015}
{McDermid} R.~M.,  et~al., 2015, \mn@doi [\mnras] {10.1093/mnras/stv105}, \href
  {http://adsabs.harvard.edu/abs/2015MNRAS.448.3484M} {448, 3484}

\bibitem[\protect\citeauthoryear{{Mitzkus}, {Cappellari}  \&
  {Walcher}}{{Mitzkus} et~al.}{2016}]{Mitzkus2016}
{Mitzkus} M.,  {Cappellari} M.,   {Walcher} C.~J.,  2016, preprint, \href
  {http://adsabs.harvard.edu/abs/2016arXiv161004516M} {} (\mn@eprint {arXiv}
  {1610.04516})

\bibitem[\protect\citeauthoryear{Mor{\'e}, Garbow  \& Hillstrom}{Mor{\'e}
  et~al.}{1980}]{More1980minpack}
Mor{\'e} J.,  Garbow B.,   Hillstrom K.,  1980, User guide for MINPACK-1.
Argonne National Laboratory Argonne, IL, \url
  {http://cds.cern.ch/record/126569}

\bibitem[\protect\citeauthoryear{{Morelli}, {Calvi}, {Masetti}, {Parisi},
  {Landi}, {Maiorano}, {Minniti}  \& {Galaz}}{{Morelli}
  et~al.}{2013}]{Morelli2013}
{Morelli} L.,  {Calvi} V.,  {Masetti} N.,  {Parisi} P.,  {Landi} R.,
  {Maiorano} E.,  {Minniti} D.,   {Galaz} G.,  2013, \mn@doi [\aap]
  {10.1051/0004-6361/201321733}, \href
  {http://adsabs.harvard.edu/abs/2013A%26A...556A.135M} {556, A135}

\bibitem[\protect\citeauthoryear{{Morelli}, {Corsini}, {Pizzella}, {Dalla
  Bont{\`a}}, {Coccato}  \& {M{\'e}ndez-Abreu}}{{Morelli}
  et~al.}{2015}]{Morelli2015}
{Morelli} L.,  {Corsini} E.~M.,  {Pizzella} A.,  {Dalla Bont{\`a}} E.,
  {Coccato} L.,   {M{\'e}ndez-Abreu} J.,  2015, \mn@doi [\mnras]
  {10.1093/mnras/stv1357}, \href
  {http://adsabs.harvard.edu/abs/2015MNRAS.452.1128M} {452, 1128}

\bibitem[\protect\citeauthoryear{{Munari}, {Sordo}, {Castelli}  \&
  {Zwitter}}{{Munari} et~al.}{2005}]{Munari2005}
{Munari} U.,  {Sordo} R.,  {Castelli} F.,   {Zwitter} T.,  2005, \aap, 442,
  1127

\bibitem[\protect\citeauthoryear{{Naab} et~al.,}{{Naab}
  et~al.}{2014}]{Naab2014}
{Naab} T.,  et~al., 2014, \mn@doi [\mnras] {10.1093/mnras/stt1919}, \href
  {http://adsabs.harvard.edu/abs/2014MNRAS.444.3357N} {444, 3357}

\bibitem[\protect\citeauthoryear{Nocedal \& Wright}{Nocedal \&
  Wright}{2006}]{nocedal2006numerical}
Nocedal J.,  Wright S.,  2006, Numerical Optimization.
Springer Series in Operations Research and Financial Engineering, Springer, New
  York, \url {https://books.google.com/books?id=VbHYoSyelFcC}

\bibitem[\protect\citeauthoryear{{Ocvirk}, {Pichon}, {Lan{\c c}on}  \&
  {Thi{\'e}baut}}{{Ocvirk} et~al.}{2006}]{Ocvirk2006}
{Ocvirk} P.,  {Pichon} C.,  {Lan{\c c}on} A.,   {Thi{\'e}baut} E.,  2006,
  \mn@doi [\mnras] {10.1111/j.1365-2966.2005.09182.x}, \href
  {http://adsabs.harvard.edu/abs/2006MNRAS.365...46O} {365, 46}

\bibitem[\protect\citeauthoryear{{Oh}, {Sarzi}, {Schawinski}  \& {Yi}}{{Oh}
  et~al.}{2011}]{Oh2011}
{Oh} K.,  {Sarzi} M.,  {Schawinski} K.,   {Yi} S.~K.,  2011, \mn@doi [\apjs]
  {10.1088/0067-0049/195/2/13}, \href
  {http://adsabs.harvard.edu/abs/2011ApJS..195...13O} {195, 13}

\bibitem[\protect\citeauthoryear{Oliphant}{Oliphant}{2007}]{Numpy2007}
Oliphant T.~E.,  2007, \mn@doi [Computing in Science \& Engineering]
  {10.1109/MCSE.2007.58}, 9, 10

\bibitem[\protect\citeauthoryear{{Onodera} et~al.,}{{Onodera}
  et~al.}{2012}]{Onodera2012}
{Onodera} M.,  et~al., 2012, \mn@doi [\apj] {10.1088/0004-637X/755/1/26}, \href
  {http://adsabs.harvard.edu/abs/2012ApJ...755...26O} {755, 26}

\bibitem[\protect\citeauthoryear{{Press}, {Teukolsky}, {Vetterling}  \&
  {Flannery}}{{Press} et~al.}{2007}]{Press2007}
{Press} W.~H.,  {Teukolsky} S.~A.,  {Vetterling} W.~T.,   {Flannery} B.~P.,
  2007, Numerical recipes: The art of scientific computing, 3rd edn.
Cambridge Univ. Press, Cambridge, \url
  {https://books.google.com/books?id=1aAOdzK3FegC}

\bibitem[\protect\citeauthoryear{{Prugniel} \& {Soubiran}}{{Prugniel} \&
  {Soubiran}}{2001}]{Prugniel2001}
{Prugniel} P.,  {Soubiran} C.,  2001, \mn@doi [\aap]
  {10.1051/0004-6361:20010163}, \href
  {http://adsabs.harvard.edu/abs/2001A%26A...369.1048P} {369, 1048}

\bibitem[\protect\citeauthoryear{{Rix} \& {White}}{{Rix} \&
  {White}}{1992}]{Rix1992losvd}
{Rix} H.-W.,  {White} S.~D.~M.,  1992, \mnras, \href
  {http://ads.ari.uni-heidelberg.de/abs/1992MNRAS.254..389R} {254, 389}

\bibitem[\protect\citeauthoryear{{SDSS Collaboration}}{{SDSS
  Collaboration}}{2016}]{sdss2016}
{SDSS Collaboration} 2016, preprint, \href
  {http://adsabs.harvard.edu/abs/2016arXiv160802013S} {} (\mn@eprint {arXiv}
  {1608.02013})

\bibitem[\protect\citeauthoryear{{S{\'a}nchez-Bl{\'a}zquez}
  et~al.,}{{S{\'a}nchez-Bl{\'a}zquez} et~al.}{2006}]{Sanchez-Blazquez2006}
{S{\'a}nchez-Bl{\'a}zquez} P.,  et~al., 2006, \mn@doi [\mnras]
  {10.1111/j.1365-2966.2006.10699.x}, \href
  {http://adsabs.harvard.edu/abs/2006MNRAS.371..703S} {371, 703}

\bibitem[\protect\citeauthoryear{{S{\'a}nchez} et~al.,}{{S{\'a}nchez}
  et~al.}{2012}]{Sanchez2012}
{S{\'a}nchez} S.~F.,  et~al., 2012, \mn@doi [\aap]
  {10.1051/0004-6361/201117353}, \href
  {http://adsabs.harvard.edu/abs/2012A%26A...538A...8S} {538, A8}

\bibitem[\protect\citeauthoryear{{Sargent}, {Schechter}, {Boksenberg}  \&
  {Shortridge}}{{Sargent} et~al.}{1977}]{Sargent1977}
{Sargent} W.~L.~W.,  {Schechter} P.~L.,  {Boksenberg} A.,   {Shortridge} K.,
  1977, \mn@doi [\apj] {10.1086/155052}, \href
  {http://adsabs.harvard.edu/abs/1977ApJ...212..326S} {212, 326}

\bibitem[\protect\citeauthoryear{{Sarzi}, {Falc{\'o}n-Barroso}, {Davies}  \&
  {et al.}}{{Sarzi} et~al.}{2006}]{Sarzi2006}
{Sarzi} M.,  {Falc{\'o}n-Barroso} J.,  {Davies} R.~L.,   {et al.} 2006, \mn@doi
  [\mnras] {10.1111/j.1365-2966.2005.09839.x}, \href
  {http://adsabs.harvard.edu/abs/2006MNRAS.366.1151S} {366, 1151}

\bibitem[\protect\citeauthoryear{{Schechter} \& {Gunn}}{{Schechter} \&
  {Gunn}}{1979}]{Schechter1979}
{Schechter} P.~L.,  {Gunn} J.~E.,  1979, \mn@doi [\apj] {10.1086/156978}, \href
  {http://adsabs.harvard.edu/abs/1979ApJ...229..472S} {229, 472}

\bibitem[\protect\citeauthoryear{Schiff}{Schiff}{1968}]{schiff1968quantum}
Schiff L.~I.,  1968, Quantum Mechanics 3rd ed..
McGraw-Hill, New York

\bibitem[\protect\citeauthoryear{{Scott} et~al.,}{{Scott}
  et~al.}{2015}]{Scott2015}
{Scott} N.,  et~al., 2015, \mn@doi [\mnras] {10.1093/mnras/stv1127}, \href
  {http://adsabs.harvard.edu/abs/2015MNRAS.451.2723S} {451, 2723}

\bibitem[\protect\citeauthoryear{{Seth} et~al.,}{{Seth}
  et~al.}{2014}]{Seth2014}
{Seth} A.~C.,  et~al., 2014, \mn@doi [\nat] {10.1038/nature13762}, \href
  {http://adsabs.harvard.edu/abs/2014Natur.513..398S} {513, 398}

\bibitem[\protect\citeauthoryear{{Shetty} \& {Cappellari}}{{Shetty} \&
  {Cappellari}}{2015}]{Shetty2015}
{Shetty} S.,  {Cappellari} M.,  2015, \mn@doi [\mnras] {10.1093/mnras/stv1948},
  \href {http://adsabs.harvard.edu/abs/2015MNRAS.454.1332S} {454, 1332}

\bibitem[\protect\citeauthoryear{{Storey} \& {Hummer}}{{Storey} \&
  {Hummer}}{1995}]{Storey1995}
{Storey} P.~J.,  {Hummer} D.~G.,  1995, \mn@doi [\mnras]
  {10.1093/mnras/272.1.41}, \href
  {http://adsabs.harvard.edu/abs/1995MNRAS.272...41S} {272, 41}

\bibitem[\protect\citeauthoryear{Swarztrauber}{Swarztrauber}{1982}]{fftpack}
Swarztrauber P.~N.,  1982, in {Rodrigue} G.,  ed., Parallel Computations.
  Academic Press, New York, pp 51--83,
  \mn@doi{10.1016/B978-0-12-592101-5.50007-5}

\bibitem[\protect\citeauthoryear{Tabor, Merrifield, Arag\'on-Salamanca,
  Bamford, Cappellari  \& Johnston}{Tabor et~al.}{2016}]{Tabor2016}
Tabor M.,  Merrifield M.,  Arag\'on-Salamanca A.,  Bamford S.~P.,  Cappellari
  M.,   Johnston E.,  2016, submitted to \mnras

\bibitem[\protect\citeauthoryear{{Thomas} et~al.,}{{Thomas}
  et~al.}{2013}]{Thomas2013boss}
{Thomas} D.,  et~al., 2013, \mn@doi [\mnras] {10.1093/mnras/stt261}, \href
  {http://adsabs.harvard.edu/abs/2013MNRAS.431.1383T} {431, 1383}

\bibitem[\protect\citeauthoryear{Tikhonov \& Arsenin}{Tikhonov \&
  Arsenin}{1977}]{tikhonov1977regul}
Tikhonov A.~N.,  Arsenin V.~Y.,  1977, Solutions of ill-posed problems.
Wiley, New York

\bibitem[\protect\citeauthoryear{{Tojeiro}, {Heavens}, {Jimenez}  \&
  {Panter}}{{Tojeiro} et~al.}{2007}]{Tojeiro2007}
{Tojeiro} R.,  {Heavens} A.~F.,  {Jimenez} R.,   {Panter} B.,  2007, \mn@doi
  [\mnras] {10.1111/j.1365-2966.2007.12323.x}, \href
  {http://adsabs.harvard.edu/abs/2007MNRAS.381.1252T} {381, 1252}

\bibitem[\protect\citeauthoryear{{Tonry} \& {Davis}}{{Tonry} \&
  {Davis}}{1979}]{Tonry1979}
{Tonry} J.,  {Davis} M.,  1979, \mn@doi [\aj] {10.1086/112569}, \href
  {http://adsabs.harvard.edu/abs/1979AJ.....84.1511T} {84, 1511}

\bibitem[\protect\citeauthoryear{{Valdes}, {Gupta}, {Rose}, {Singh}  \&
  {Bell}}{{Valdes} et~al.}{2004}]{Valdes2004}
{Valdes} F.,  {Gupta} R.,  {Rose} J.~A.,  {Singh} H.~P.,   {Bell} D.~J.,  2004,
  \mn@doi [\apjs] {10.1086/386343}, \href
  {http://adsabs.harvard.edu/abs/2004ApJS..152..251V} {152, 251}

\bibitem[\protect\citeauthoryear{{Vazdekis}, {S{\'a}nchez-Bl{\'a}zquez},
  {Falc{\'o}n-Barroso}, {Cenarro}, {Beasley}, {Cardiel}, {Gorgas}  \&
  {Peletier}}{{Vazdekis} et~al.}{2010}]{Vazdekis2010}
{Vazdekis} A.,  {S{\'a}nchez-Bl{\'a}zquez} P.,  {Falc{\'o}n-Barroso} J.,
  {Cenarro} A.~J.,  {Beasley} M.~A.,  {Cardiel} N.,  {Gorgas} J.,   {Peletier}
  R.~F.,  2010, \mn@doi [\mnras] {10.1111/j.1365-2966.2010.16407.x}, \href
  {http://adsabs.harvard.edu/abs/2010MNRAS.404.1639V} {404, 1639}

\bibitem[\protect\citeauthoryear{Voglis \& Lagaris}{Voglis \&
  Lagaris}{2004}]{voglis2004dogleg}
Voglis C.,  Lagaris I.,  2004, in WSEAS International Conference on Applied
  Mathematics. Corfu, Greece, \url
  {http://www.wseas.us/e-library/conferences/corfu2004/papers/488-317.pdf}

\bibitem[\protect\citeauthoryear{{Walsh}, {van den Bosch}, {Gebhardt},
  {Y{\i}ld{\i}r{\i}m}, {Richstone}, {G{\"u}ltekin}  \& {Husemann}}{{Walsh}
  et~al.}{2016}]{Walsh2016}
{Walsh} J.~L.,  {van den Bosch} R.~C.~E.,  {Gebhardt} K.,  {Y{\i}ld{\i}r{\i}m}
  A.,  {Richstone} D.~O.,  {G{\"u}ltekin} K.,   {Husemann} B.,  2016, \mn@doi
  [\apj] {10.3847/0004-637X/817/1/2}, \href
  {http://adsabs.harvard.edu/abs/2016ApJ...817....2W} {817, 2}

\bibitem[\protect\citeauthoryear{{Weijmans} et~al.,}{{Weijmans}
  et~al.}{2009}]{Weijmans2009}
{Weijmans} A.-M.,  et~al., 2009, \mn@doi [\mnras]
  {10.1111/j.1365-2966.2009.15134.x}, \href
  {http://adsabs.harvard.edu/abs/2009MNRAS.398..561W} {398, 561}

\bibitem[\protect\citeauthoryear{{Westfall}, {Bershady}  \&
  {Verheijen}}{{Westfall} et~al.}{2011}]{Westfall2011}
{Westfall} K.~B.,  {Bershady} M.~A.,   {Verheijen} M.~A.~W.,  2011, \mn@doi
  [\apjs] {10.1088/0067-0049/193/1/21}, \href
  {http://adsabs.harvard.edu/abs/2011ApJS..193...21W} {193, 21}

\bibitem[\protect\citeauthoryear{Wolfram}{Wolfram}{2003}]{wolfram2003mathematica}
Wolfram S.,  2003, The Mathematica Book, 5th ed..
Wolfram Media, Champaign, IL, \url {http://reference.wolfram.com/}

\bibitem[\protect\citeauthoryear{{Yan} et~al.,}{{Yan} et~al.}{2016}]{Yan2016}
{Yan} R.,  et~al., 2016, preprint, \href
  {http://adsabs.harvard.edu/abs/2016arXiv160708613Y} {} (\mn@eprint {arXiv}
  {1607.08613})

\bibitem[\protect\citeauthoryear{{de Zeeuw} et~al.,}{{de Zeeuw}
  et~al.}{2002}]{deZeeuw2002}
{de Zeeuw} P.~T.,  et~al., 2002, \mn@doi [\mnras]
  {10.1046/j.1365-8711.2002.05059.x}, \href
  {http://adsabs.harvard.edu/abs/2002MNRAS.329..513D} {329, 513}

\bibitem[\protect\citeauthoryear{{van de Sande} et~al.,}{{van de Sande}
  et~al.}{2016}]{vanDeSande2016}
{van de Sande} J.,  et~al., 2016, submitted

\bibitem[\protect\citeauthoryear{{van der Marel}}{{van der
  Marel}}{1994}]{vanDerMarel1994m87}
{van der Marel} R.~P.,  1994, \mnras, \href
  {http://adsabs.harvard.edu/abs/1994MNRAS.270..271V} {270, 271}

\bibitem[\protect\citeauthoryear{{van der Marel} \& {Franx}}{{van der Marel} \&
  {Franx}}{1993}]{vanDerMarel93}
{van der Marel} R.~P.,  {Franx} M.,  1993, \mn@doi [\apj] {10.1086/172534},
  \href {http://adsabs.harvard.edu/abs/1993ApJ...407..525V} {407, 525}

\bibitem[\protect\citeauthoryear{{van der Marel}, {Rix}, {Carter}, {Franx},
  {White}  \& {de Zeeuw}}{{van der Marel} et~al.}{1994}]{vanderMarel1994}
{van der Marel} R.~P.,  {Rix} H.~W.,  {Carter} D.,  {Franx} M.,  {White}
  S.~D.~M.,   {de Zeeuw} T.,  1994, \mnras, \href
  {http://ads.ari.uni-heidelberg.de/abs/1994MNRAS.268..521V} {268, 521}

\makeatother
\end{thebibliography}

\bsp    
\label{lastpage}
\end{document}